\documentclass[preprint,12pt,authoryear]{elsarticle}

\usepackage{amsmath,amsfonts,amssymb,amsthm,subfig,subfloat}
\usepackage{sublabel}
\usepackage[dvipsnames]{xcolor}
\usepackage{url}
\usepackage{fancyhdr}
\usepackage{alltt}
\usepackage{ifthen}
\usepackage[latin1,utf8]{inputenc}
\usepackage[T1]{fontenc}
\usepackage{times}
\usepackage{graphicx}
\usepackage{array}
\usepackage[pdftex,colorlinks]{hyperref}
\usepackage{verbatim}
\usepackage{fancyvrb}
\usepackage{placeins}
\usepackage{tabularx}

\graphicspath{{../figs/}}  

\newcommand{\Nu}{\mbox{\it Nu}}

\newcommand{\bnabla}{\mbox{\boldmath $\nabla$}}
\renewcommand{\vec}[1]{\mbox{\boldmath $#1$}}
\newcommand{\vechat}[1]{{\skew3\hat{\vec{#1}}}}

\newcommand{\pd}[1]{\partial_{#1}}

\newcommand{\curl}{\bnabla \wedge}

\newcommand{\beq}{\begin{equation}}
\newcommand{\eeq}{\end{equation}}

\begin{document}

\begin{frontmatter}



\title{Can a convecting magma ocean offer a solution to the puzzling case of core convection in early earth?}


\author[add1]{Urmi Dutta\corref{cor1}}
\ead{urmidutta@pup.ac.in}
\cortext[cor1]{Corresponding author}
\author[add2]{Chris J. Davies}
\author[add3]{Ashley P.  Willis\corref{cor1}}
\ead{a.p.willis@sheffield.ac.uk}

\address[add1]{Department of Geology, Patna University, Patna- 800005, India}
\address[add2]{School of Earth and the Environment, University of Leeds, Leeds LS2 9JT, United Kingdom}
\address[add3]{Applied Mathematics, School of Mathematical and Physical Sciences, University of Sheffield, Sheffield S3 7RH, United Kingdom}

\begin{abstract}
Convective flow in Earth's iron-rich liquid core drives self-sustained dynamo action, generating Earth's magnetic field, which is strongest among all terrestrial planets of the solar system. Rock records show that this magnetic field has been operative in Earth for at least 3.4 billion years (b.y). However, advanced high pressure experiments have revised the value of the thermal conductivity of the outer core, which implies an age for the inner core of less than 1 b.y., 
when compositional convection begins. 
This creates a puzzle, with a gap between 
the observations of an early magnetic field on Earth and the young inner core. 
Previous work has suggested that the pre-inner core dynamo could have been generated in a magma ocean (MO) at the base of the mantle; however, the fluid dynamics of this scenario have received little attention. Here we numerically model the non-magnetic rotating flow in a MO above a
convectively stable core 
in a configuration representing the pre-inner core days of Earth's evolution.
Simulations here explore the importance of several dimensionless parameters on 
coupled core-MO convection
-- the Rayleigh number, 
the ocean/core 
thermal diffusivity ratio, 
thermal expansion coefficient ratio,  
viscosity ratio,  
and layer thickness ratio. 
It is found that the MO can easily drive a flow of comparable magnitude in the core,
and an approximately linear relationship is observed between the ratio of root-mean-square velocities in the core and the ocean, $(u_c^{RMS}/u_o^{RMS})$, and $(\Nu_o-1)$, where $\Nu_o$ is the Nusselt number for the MO, for the $\Nu_o$ of order 1 to 10 considered.
Radial and azimuthal components of the core flow are of similar magnitude, so that, with comparable toroidal and poloidal components, 
we speculate that
the MO-driven core flow could drive an early dynamo.

\end{abstract}


\end{frontmatter}


\section{Introduction} 

One of the most important events in the 4.5 b.y.\ of Earth's history is the 
formation of the core.  Differentiation of the metal core 
from the silicate mantle took place in a global magma ocean (MO).
The MO itself is believed to have been created when the primitive undifferentiated Earth suffered large-scale melting \citep{stixrude2009thermodynamics} due to
a moon-forming impact event \citep{rubie2007formation}.
Soon after formation of the MO, gravitational separation caused the heavier metal component to sink towards the center to form the core. The remaining MO then consisted mostly of molten silicate.

The huge gravitational potential energy of the core is slowly released
as heat.  This cooling ultimately resulted in the crystallization in the metallic 
inner core. Subsequent release of lighter elements from the inner core surface  triggered compositional convection in the liquid outer core. However many workers have argued in favour of a much younger inner core. \cite{labrosse2001age} worked on the analytical and computational models of heat balance within earth's core with and without radioactive elements. It was found that without any radioactive element inner core did not start crystallizing before 1 b.y. making it much younger
than the primitive core separation event. Later works by \cite{pozzo2013transport}, \cite{davies2015constraints}, \cite{labrosse2015thermal}, and \cite{nimmo20159} 
used updated (higher) values of the core's thermal conductivity
for iron, suggesting the inner core to be not much older than  1 b.y.\ and more likely to be as young as $\approx$ 0.5 b.y.  This creates a new problem for the energy source for Earth's early magnetic field because only thermal convection in highly conductive core may not provide enough power to sustain the geodynamo \citep{driscoll2023new}. \cite{olson2013new} labelled this conflicting case as the ``new core paradox'' and hinted that the explanation may be found when we think beyond the conventional theory of core evolution. Several works have considered alternative mechanisms to drive core-convection 
before inner core formation,
where suggestions have included colder subducting plates and induced lateral heat variation at the CMB \citep{olson2016mantle}, Mg, SiO$_2$ or Si precipitation in core \citep{o2016powering, badro2016early, hirose2017crystallization, wilson2022powering}. \cite{driscoll2023new} derived a number of `successful' and `unsuccessful' solutions to tackle the new core paradox and identified one promising scenario where the dynamo could have acted throughout the Earth's history. The high CMB temperature in this case might have helped to keep the mantle above its solidus, preserving an MO longer by delaying the crystallization.

The thickness of the MO depends on the amount of initial melting caused by the impact as well as the nature of the crystallization within the silicate MO. Most MO studies have focussed on this crystallisation \citep{elkins2003magma, labrosse2007crystallizing, solomatov2015magma, stixrude2009thermodynamics, coltice2011crystallization, nomura2011spin}. Many considered that the process starts from an intermediate depth where the 
adiabat had a steeper slope than the strongly curved liquidus
 favoring presence of a partially molten layer in the lower mantle. This residual melt layer, stable at the core-mantle boundary, has been identified as basal magma ocean (BMO) in literatures \citep{labrosse2007crystallizing, stixrude2009thermodynamics, nomura2011spin}.
It has been estimated \citep{labrosse2007crystallizing}, that due to the crystallisation within the mantle, eventually a 1000 km thick BMO formed at the top of the core. The presence of a basal magma ocean also gained support from shock experiments on deep melt layers \citep{mosenfelder2007thermodynamic}.  On the other hand, some workers have suggested that the magma ocean froze from the bottom upwards \citep{thomas2012multi, ballmer2017reconciling} which would make the formation
of a basal MO unlikely.
However, \cite{nomura2011spin} found that at deep mantle conditions, a spin crossover of iron (from high-spin to low-spin) is responsible for its fractionation to the silicate melt. This makes the melt denser and favours the formation of the basal magma ocean, irrespective of the depth at which the crystallization started. 
This idea gained support from the thermodynamic analysis of \cite{stixrude2009thermodynamics}. Their first principles molecular dynamic simulations indicated
 that the liquid is denser than the solid phase at 4000\,K and pressures greater than 130\,GPa. Also, high pressure studies
corresponding to a deep MO \citep{murakami2011evidence} suggest that silicate melts are indeed stable at the core mantle boundary (CMB). \cite{labrosse2015fractional} and  \cite{caracas2019melt} have argued that a BMO is inevitable regardless of how the mantle froze. Numerical models of MO evolution by \cite{boukare2025solidification}, which also takes account of the geochemical and petrological data of the lower mantle, produce a BMO even with the least favorable conditions.

  A peridotic, terrestrial MO has high temperature, low viscosity and thus high Rayleigh number \citep{huang2024low} implying that it would convect vigorously under a turbulent flow regime. Some studies have considered multiphase modelling of MO flows and also their geochemical consequences \citep{boukare2023lava, boukare2025solidification} for Earth and  exoplanets. \cite{morison2019timescale}, \cite{agrusta2020mantle}, and \cite{labrosse2024solid} have modeled surface and basal MO convections, having interaction with solid-state flow in the mantle. Their works have considered the thermodynamic influence of the phase change within the MO due to any crystallization or melting taking place along the mantle-MO interface. The lower phase change number ($\Phi$) suggests greater interaction between the MO and mantle. 
  
  Our current study focuses on the mechanical interaction between the MO and the core. Both layers have a long viscous timescale and thus high value of $\Phi$, implying very little phase change at the 
  core-mantle boundary.  The MO in our models represent the fully molten, single phase, heavier residue seggregated at the bottom during MO crystallization. 
  
A potential resolution to the dynamo paradox is a dynamo within the BMO.
  However, whether there is a prospect of MO convection contributing to dynamo action depends its electrical conductivity. First-principles molecular dynamics simulation \citep{scipioni2017electrical} showed that at extremely high pressure, temperature, conditions for an Earth-sized planet, the electrical conductivity value is 5700 S.m$^{-1}$. They added that, not only might Earth have had an ancient magma ocean generated magnetic field, but that other super-earth exoplanets with a molten silicate interior are also capable of generating magnetic field. Whereas this was a purely molecular dynamic study, \cite{laneuville2018crystallization} thermochemically modelled a coupled MO-Core system for Earth. They suggested that if the MO remained stably stratified, the core dynamo might act. However, when the MO starts to convect and become well mixed, the core dynamo stops. \cite{blanc2020thermal} showed that it was thermodynamically possible to sustain an early dynamo within a downward crystallizing BMO overlying the liquid outer core. The BMO-powered dynamos in their study were strongly influenced by the incompatibility of iron in silicate melt at near-CMB conditions. Their model is also applicable to study the evolution of earth-like-planets or other exo-planets.
  
As researchers have started to focus on the spatial convective dynamics of the MO in last decade, there is still a long way to catch up with the huge repository of studies on core flows.
In this paper we aim to understand better the range of types of convective
flows that could be generated in the magma ocean. 
While the MO covered the core, it is expected that the inner core
had not yet formed and thus the coupling might have affected the core evolution.
We model the core and MO fluid layers as concentric spherical shells 
in a rotating frame, and study the effects of their differing parameter 
ratios on the thermal and viscous coupling,  the spatial dynamics
and their affect on core cooling that this induces.  We do not yet
involve a magnetic field.
Given the discussion above, we 
suppose that secular cooling drives the flow, 
rather than in an internal heat source, such as radioactive decay.
Mathematically, however, it is convenient to recast the cooling 
such that appears as internal source term in the equation governing
the temperature.  We identify
this term with the cooling, and seek to identify the conditions
under which it can generate  
flows with coupling between
the layers.
The spatio-temporal complexity of the flows might be considered a simple proxy for its capacity to act as a dynamo.

\section{Model and numerical method}

\subsection{Material properties for the model}

The fluid model we are considering will have two concentric layers: 1. Top- Magma ocean(o), 2. Bottom- core (c).

The initial composition of the magma ocean is suggested to be fully molten silicates \citep{rubie2003mechanisms} whereas the core is mainly composed of liquid metal e.g. Ni, Fe \citep{mcdonough1995composition} with light elements yet to be fractionated. Over the course of the crystallization of the MO, it is believed to be consisting of of two phases-crystal and melt \citep{morison2019timescale, agrusta2020mantle, labrosse2024solid}. However, according to the study by \citep{caracas2019melt}, the crystallization history and melt-crystal equilibrium within magma ocean is rather complex. Among different crystallization scenarios, they favored the case where the batch crystallization of bridgmanite caused compaction induced downward melt migration above the density crossover depth (DCD). This resulted in the segregation of the BMO of Fe-rich dense melt which is essentially a single phase separated from compacted Fe-poor cumulate mush. The physical properties of the liquid outer core has long been a topic of research in the field of geophysics, high-pressure, high-temperature experiments and in mineral physics. According to the PREM model, {\it P-} and \textit{S-}wave velocity predicts the density of the outer core ($\rho_c$) to be $\sim$ $10^4$ Kg/m$^3$ (after \citep{dziewonski1981preliminary, masters1995seismic}. The \textit{Ab initio} calculations by \cite{pozzo2013transport} yielded values for density, viscosity, thermal conductivity and specific heat of Earth's core. From those values we can estimate the dynamic viscosity ($\mu_c$) and the thermal diffusivity of outer core ($\kappa_c$) to be 10$^{-2}$-10$^{-3}$ Pa.s and $10^{-5}$ m$^2$/s respectively. On the other hand there are very few studies on the possible material properties of the magma ocean that existed at the early days of core formation. 
\cite{solomatov2015magma} listed the density of the magma ocean($\rho_o$) as 4x $10^3$ Kg/m$^3$.
\begin{table} [!h]
\hspace{-20mm}
\begin{tabular}{||p{0.37\linewidth}|p{0.14\linewidth}|c|p{0.4\linewidth}||}
\hline
\hline
\textbf{Parameters}		&\textbf{Symbol}   		&  \textbf{Value} 	& \textbf{Reference}\\
\hline
Density of the Magma ocean 			&$\rho_o$	 &4 $\times$ 10$^3$ Kg/m$^3$	& \cite{solomatov2015magma}	\\
\hline
Density of the Outer core 				& $\rho_c$			&$10^4$ Kg/m$^3$ & PREM e.g.\cite{mcdonough2003compositional}		\\
\hline
Density Ratio			& $\rho_{o/c}$	&	$2/5$	& \\
\hline
Viscosity of the Magma ocean 	& $\mu_o$		&	10$^{-2}$-10	Pa\,s & \cite{ichikawa2010direct}\\
\hline
Viscosity of the Outer core & $\mu_c$ & 10$^{-2}$-10$^{-3}$ Pa\,s	& \cite{pozzo2013transport} \\
\hline
Viscosity ratio			& $\mu_{o/c}$		& 1-100		 & \\
\hline
Thermal diffusivity of the Magma ocean 			&$\kappa_o$	&$10^{-6}$ m$^2$/s & \cite{lebrun2013thermal}	\\
\hline
Thermal diffusivity of the Outer core 		&$\kappa_c$	&$10^{-5}$ m$^2$/s & \cite{davies2015constraints}$^*$	\\
\hline
Thermal diffusivity Ratio			&$\kappa_{o/c}$		 &$0.1$	& \\
\hline
Thermal expansion coefficient of the Magma ocean 		&$\alpha_o$	&5 $\times$ $10^{-5}$ /K & \cite{solomatov2015magma}	\\

\hline
Thermal expansion coefficient of the Outer core 	&$\alpha_c$	&1 $\times$ $10^{-5}$ /K & \cite{gubbins2003can}	\\
\hline
Thermal expansion coefficient Ratio			&$\alpha_{o/c}$	 &$\approx$ 5	& \\
\hline
Thickness of the Magma ocean 			&$\delta_o$	 & 1000 km		& \cite{labrosse2007crystallizing}\\
\hline
Thickness of the Outer core 				& $\delta_c$		&3480 km  & PREM e.g.\cite{mcdonough2003compositional}		\\
\hline
Thickness Ratio                         &$\delta_{o/c+o}$		&$\approx$ 0.22	 &	\\
\hline
       \end{tabular}\
           \caption{Dimension and material properties of the Magma ocean model.
           $x_{o/c}$ denotes $x_o/x_c$ and
           $\delta_{o/c+o}=\delta_o/(\delta_c+\delta_o)$. 
           ~~
           $^*$ $\kappa=k/(\rho\,C_p)$.
}
\label{table1}
\end{table}
The numerical studies on metal-silicate interaction in MO \citep{ichikawa2010direct} and phase separation during MO crystallisation \citep{boukare2017modeling} estimated the magma ocean viscosity ($\mu_o$) to be varied in the range of  10$^{-2}$-10 Pa.s and 10$^{-1}$-10$^{3}$ Pa.s respectively. This value later gained support from the first-principles molecular dynamics simulations done by \cite{Karki740}. \cite{lebrun2013thermal}, in their parametric study on magma ocean-atmosphere interaction considered the thermal diffusivity of magma ocean ($\kappa_o$) in the order of $10^{-6}$ m$^2$/s.

The thickness of the basal magma ocean ($\delta_o$) is assumed to be 1000 km and the radius of the core is 3480 km (PREM e.g.\cite{mcdonough2003compositional}). 
Prior to inner core formation, the outer core radius
($\delta_c$) will match the total core radius i.e. $\approx$ 3480 km.

\subsection{Governing equations and method}
\label{subsec:nummethod}

In spherical coordinates $(r,\theta,\phi)$, 
the dimensional governing equations for the 
velocity $\vec{u}$ and temperature $\Theta$ 
in a frame with rotation rate $\Omega$ are
\begin{eqnarray}
  \label{eq:acc_dimensional}
  \partial_t\vec{u}+\vec{u}\cdot\bnabla\vec{u} &=& \nu\,\bnabla^2\vec{u}+
\alpha \,g(r)\, \Theta \vechat{r} - 2\,\Omega\,\vechat{z}\times\vec{u} - 
\frac{1}{\rho}\bnabla p\, ,
\\
\bnabla\cdot\vec{u} &=& 0\, ,
\\
   \partial_t\Theta+\vec{u}\cdot\bnabla \Theta &=& \kappa\nabla^2\Theta + \varepsilon \, ,
\end{eqnarray}
where $\varepsilon$ is an internal heat source, equivalent to an overall cooling rate across the core and MO, and $p$ is the pressure.  We denote the radius to
the core-ocean interface by $r_l$ and to the upper surface of the MO by
$r_o$ respectively.  For the physical parameters, 
$\nu$, $\alpha$, $\rho$ and $\kappa$, 
let subscripts $c$ and $o$ denote the core and ocean values, e.g.\ $\nu_c$ and $\nu_o$.
The radially-dependent gravity  
takes the form
\begin{equation}
  g(r) = \left\{  \begin{array}{ll}
  \beta\, r, & r \le r_l \\
  \beta\,(r_l^3 + (\rho_o/\rho_c)(r^3-r_l^3))/r^2, &
  r_l < r < r_o.
  \end{array} \right.
\end{equation}
where $\beta$ is a constant such that $g(r_o)=g_o$.  After the linear increase
within the core, for 
$\rho_o/\rho_c=0.4$, this drops 
at a similar rate over a MO with 
$\delta_o/(\delta_c+\delta_o)=0.2$, but
flattens out over a thicker MO.

For non-dimensionalisation we use physical parameter values associated with the core, and 
for the length scale use
$d\approx r_o$ (see paragraph on numerical implementation below).
Gravity is scaled by the value at the surface of the MO, $g_o$.
Using the viscous time scale $d^2/\nu_c$ 
and $\Delta \Theta=
\varepsilon\,d^2/\nu_c$ for the temperature scale, we 
arrive at the dimensionless equations, e.g.~\cite{ChEtal01},
\begin{eqnarray}
  \label{eq:mom}
  E\,(\partial_t+\vec{u}\cdot\bnabla-\tilde{\nu}\bnabla^2)\,\vec{u}&=&
\tilde{\alpha}\,Ra \,g(r)\, \Theta \,\vechat{r} 
- 2\,\vechat{z}\times\vec{u} 
-\frac{1}{\tilde{\rho}}\bnabla p\, , \\
   \label{eq:temp}
   \partial_t\Theta+\vec{u}\cdot\bnabla \Theta&=&
	\frac{1}{Pr}\,\tilde{\kappa}\nabla^2\Theta + 1 \, ,
\end{eqnarray}
where `tilde' denotes domain-dependent parameters normalised by the values in the core, e.g.\ 
$\tilde{\nu}(r)=\nu_o/\nu_c$ for $r_l<r<r_o$ (within the MO), and 
$\tilde{\nu}(r)=1$ for $0<r<r_l$ (within the core).
The tilde-notation avoids repeating governing equations for each domain and awkward exposition of boundary conditions, but is only required in this section.  We use ratios frequently throughout the rest of the text, however, and therefore
for brevity let the `$o/c$' subscript denote the ratio, e.g.\ $\nu_{o/c}=\nu_o/\nu_c$.

The Ekman number, modified Rayleigh number and Prandtl number
are given respectively by
\begin{equation}
   E = \frac{\nu_c}{\Omega\,d^2} , \quad 
	 Ra= \frac{\alpha_c\,g_o\,\Delta\Theta\,d}{\nu_c\,\Omega}, \quad
	 Pr=\frac{\nu_c}{\kappa_c} ,
\end{equation}
where $\Delta\Theta$ is the temperature scale used above.
These parameters are based on values for the core.  
For the MO, they are defined correspondingly or can be determined using the ratios via
\begin{equation}
E_\mathrm{MO}= \frac{\nu_{o/c}}{\delta_{o/c+o}^{2}}\,E\,,
\quad
Ra_\mathrm{MO}= \frac{\alpha_{o/c}\,\delta_{o/c+o}}{\nu_{o/c}}\,Ra 
\, , \quad
Pr_\mathrm{MO} = \frac{\nu_{o/c}}{\kappa_{o/c}} \, Pr \,.
\end{equation}
(Note that these values are not used in the governing equations -- the relative difference is set through the domain-dependent parameters, e.g.\ $\tilde{\nu}(r)$.  The temperature scale $\Delta\Theta$ is common to the layers, as flux from the core is driven through the MO.)
The Nusselt numbers, which measure the observed heat flux relative to that for pure conduction, are given by 
\begin{equation}
\Nu_o = \frac{\Theta_l'-\Theta_o'}{\Theta_l-\Theta_o},
\qquad
\Nu_c = \frac{\Theta_c'-\Theta_l'}{\Theta_c-\Theta_l},  
\end{equation}
where $\Theta_l$ and $\Theta_o$ denote the spatial average of $\Theta$ evaluated at the interface and the top of the MO respectively, $\Theta_c$ is the spatial average over the whole core, and primes indicate purely conductive values.

The boundary conditions for the velocity we assume are 
the no-slip condition at the upper surface of the MO ($r=r_o$) and 
the non-penetrative condition applied at the interface between the core and MO at $r=r_l$.
We also have continuity of parallel components of the velocity at the interface,
$[u_\theta]=[u_\phi]=0$, 
where $[f(r)]\equiv \lim_{\epsilon\to 0}(f(r_l+\epsilon)-f(r_l-\epsilon))$.
For the temperature field, 
the heat flux at the surface must match the internal heating, 
which implies the condition
\begin{equation}
   -\,4\pi\,r_o^2 \,\frac{1}{Pr}
\left.(\tilde{\kappa}\,\partial_r\Theta)\right|_{r_o} 
= \frac{4}{3}\pi\,r_o^3\,
   ~~\Rightarrow~~
   \left.(\partial_r\Theta)\right|_{r_o} = -\, 
   \frac{r_o\,\kappa_c}{3\,\kappa_o}\,Pr \, .
\end{equation}
Matching temperature and flux at the interface give
\begin{equation}
    [\Theta] = 0, \quad
    [\tilde{\kappa}\,\partial_r \Theta]=0.
\end{equation}

The numerical implementation of the above equations and interface conditions
required only a minor modification of the code of 
\cite{willis2007thermal}
which has been written with an inner core.  
To minimise modifications, we consider
free-slip of the fluid against a small 
isothermal
core of radius $r_i$: here, $d=r_o-r_i$
and we set $r_i/r_o=0.05$.  This numerical inner core
accounts for approximately only 0.01\% 
of the volume ($(r_i/r_o)^3=1.25\times 10^{-4}$) and has minimal influence on the dynamics.
The numerical formulation employs the
toridal-poloidal decomposition for the divergence-free
velocity field
\begin{equation}
   \vec{u} = \curl(T\vec{r}) + \curl\curl(P\vec{r})  
\end{equation}
The non-penetrative condition at $r_i,r_o$ and at the interface
$r_l$ are
$
   P = 0 \, ,
$
and at $r_o$, the no-slip conditions are
$
   T=0, ~ 
   \pd{r}P=0 \, .
$
The stress-free conditions on the inner core are
$
   \pd{r}(T/r) = 0,
   ~ 
   \pd{rr}P=0\, .
$
Velocities match at the interface,
$
   [T]=0, ~
   [\partial_r P]=0\, ,
$
and likewise stress,
$
   [\tilde{\nu}\,\partial_r T]=0, ~
   [\tilde{\nu}\,\partial_{rr} P]=0\, .
$
The code uses a finite difference
scheme in the radial dimension.  
To introduce the interface at $r_l$
we simply use one-sided differences as we approach the point at $r_l$.  
As the functions are single valued at this point, 
$[T]=[P]=[\Theta]=0$ are implicitly satisfied, and
$[\tilde{\nu}\,\partial_{rr} P]=0$ follows from $[g]=0$ 
(see \cite{willis2007thermal} equations (13)-(14)).
The remaining conditions to be imposed at the interface are 
$[\partial_r P]=[\tilde{\nu}\,\partial_r T]
=[\tilde{\kappa}\,\partial_r \Theta]=0$.
To ensure all conditions on $P$ are satisfied simultaneously,
Greens functions are required corresponding to the boundary 
conditions $g_G=\{1,0,0\},\,\{0,0,1\}$ and now additionally $\{0,1,0\}$
at $r=\{r_i,r_l,r_o\}$
(see \cite{willis2007thermal} equations (15)-(16)).

The default parameters used in this study are 
$E=5\times10^{-5}$, $Pr=1$, 
$\nu_{o/c}=\nu_o/\nu_c=25$, $\kappa_{o/c}=\kappa_o/\kappa_c=0.1$, 
$\alpha_{o/c}=\alpha_o/\alpha_c=5$,
$\rho_{o/c}=\rho_o/\rho_c=0.4$, $\delta_o/(\delta_c+\delta_o)=0.2$ and $r_i/r_o=0.05$.

For convection between two flat plates with the free-slip condition,
Rayleigh (1916) found the wavelength of a pair of rolls at onset to be
$\lambda=2\sqrt{2}\,\delta$, where $\delta$ is the depth.  
Supposing that a dimensionless depth of $0.2$ 
corresponds to a shallow magma ocean,
Rayleigh's result suggests first onset with azimuthal wavenumber 
$m\approx(2\pi/0.2)/(2\sqrt{2})\approx11$.
Although the boundary conditions are different, with no slip 
at the top and approximately a free-slip condition at the layer interface,
we find that $m=11$ is the first unstable mode for a 
critical Rayleigh number $Ra_c\approx4.7$,
although other $m$ modes are also very close to critical.  
When driven harder, higher wavenumbers are favoured.

\section{Results}

\begin{table} [!htb]
\begin{center}\
\begin{tabular}{||c|c|c|c|c||}
\hline
\hline
	$Simulation~name$ &$u_c^{RMS}$ 	&  $u_o^{RMS}$  	&   ${U_{o/c}}$ & $\Nu_o$\\

\hline
		$D$ &  4.99 & 16	& 3.2 & 3.07	\\
\hline
		$\alpha\_3$ &  3.26 &	12.02 & 3.68 & 2.76	\\
\hline
		$\alpha\_1$ & 1.49 & 6.28 & 4.19 & 1.90	\\
\hline
	$\kappa\_0.3$  & 2.84 & 11.94 & 4.2  & 1.55 
	\\
\hline
	$\kappa\_0.5$  & 1.71 & 8.53 & 4.98  & 1.10
	\\
\hline
	$\nu\_15$  & 5.73 & 18.03 & 3.14 &  3.30
	\\
\hline
	$\nu\_5$  &7.82 & 25.3 & 3.23  & 3.83
	\\
\hline
		$\delta\_0.3$  &	9.62 & 23.42	& 2.43  &  4.10 	\\
 \hline
		$\delta\_0.5$  &17.36 & 33.54	& 1.93	&   5.73\\
\hline
		$Ra\_50$   & 1.63 & 7.13	& 4.37	& 2.05 \\
\hline
		$Ra\_20$  & 0.87 & 5.10	& 5.83	& 1.44 \\ 
\hline
       \end{tabular}\
 \end{center}
           \caption{ 
           Summary of parameter sets and relative magnitudes of velocity. 
           The Ekman number, modified Rayleigh number and Prandtl number for the core
           are fixed at $Ra=200$, $E=5\times 10^{-5}$ and $Pr=1$ respectively.
           Simulation $D$ is for the default parameter set, i.e.\ $\nu_{o/c}=25$, $\kappa_{o/c}=0.1$, $\alpha_{o/c}=5$, $\rho_{o/c}=0.4$, $\delta_o/(\delta_c+\delta_o)=0.2$. 
           Each of the other simulations differ from $D$ by 
           a single modified parameter value; the simulation is named after the parameter and its new value.
           Nusselt numbers for core, $\Nu_c$, remain $\approx$1 in all cases.
                     }
\label{table2}
\end{table}

Our aim is to study the effect of dimensionless parameter ratios, the viscosity ratio ($\nu_{o/c}$), 
thermal diffusivity ratio ($\kappa_{o/c}$), thermal expansion coefficient ratio ($\alpha_{o/c}$) and layer thickness ratio ($\delta_{o/c+o}\equiv\delta_o/(\delta_c+\delta_o)$).
In particular, we are interested in the core-MO coupling, the
relative magnitude of the flow this induces,
$U_{o/c}=u_o^{RMS}/u_c^{RMS}$, 
where $RMS$ indicates the root-mean-square value over the respective domain,
the overall pattern of the flow,
and the heat transport as measured by $\Nu_o$.
Such factors
are expected to have a profound effect on the viability of the dynamo
mechanism within each layer.
A summary of parameters used in our simulations is given in Table \ref{table2}.  At these parameters, 
the core is stable to convection, see
e.g.\ Fig. \ref{fig:fig4}(c,f), where flow in the core is clearly driven purely by flow in the MO.  In e.g.\ Fig.\ \ref{fig:fig6}
it is observed that the MO-driven flow essentially doesn't affect temperature profile in the core
so that $\Nu_c\approx 1$.
For the default simulation $D$,
parameters for the 
the MO are $E_\mathrm{MO}=3.125\times 10^{-2}$, $Ra_\mathrm{MO}=8$ and $Pr_\mathrm{MO}=250$.  The `classical' Rayleigh number is given by the combination $Ra_\mathrm{MO}\,Pr_\mathrm{MO}/E_\mathrm{MO}=6.4\times 10^4$, which is found to be approximately 40 times critical in this configuration.  If we consider the temperature equation (\ref{eq:temp}) with the large $Pr_\mathrm{MO}=250$, we see that conduction in the MO is very inefficient at passing the heat flux coming from the core through the MO, leading to convective instability in the MO.

\subsection{Default model}

\label{subsec:Default_model}

Using data in Table \ref{table1},
approximate ratios for the
default model are taken to be $\nu_{o/c}=25$, $\kappa_{o/c}=0.1$, $\alpha_{o/c}=5$, 
$\rho_{o/c}=0.4$, $\delta_{o/c+o}=0.2$,
which are given alongside 
the chosen Rayleigh
number in the first row of Table \ref{table2}.
The Ekman number 
$E = 5\times 10^{-5}$
is chosen to be as small as computationally manageable,
whilst being sufficiently small to clearly see rotational effects.
Given these parameter values, the default Rayleigh number $Ra = 200$ is approximately $40$ times supercritical. The regime diagram derived by \cite{long2020scaling}, in the space of super criticality of the flow and Ekman number, demarcates regimes of weakly nonlinear, rapidly rotating, transitional and non-rotating convection in Earth-like configuration for the core. According to the parameter values, our default model ($Ra/Ra_c= 40$, $E = 5\times 10^{-5}$) lies in the transitional-regime criteria
for the core region.
Their spherical-shell transitional regime models showed sensitivity towards rotational force and were not entirely columnar in nature.

\begin{figure} 
 \begin{center}
\subfloat[]{\includegraphics[angle =0, width=0.45\textwidth]{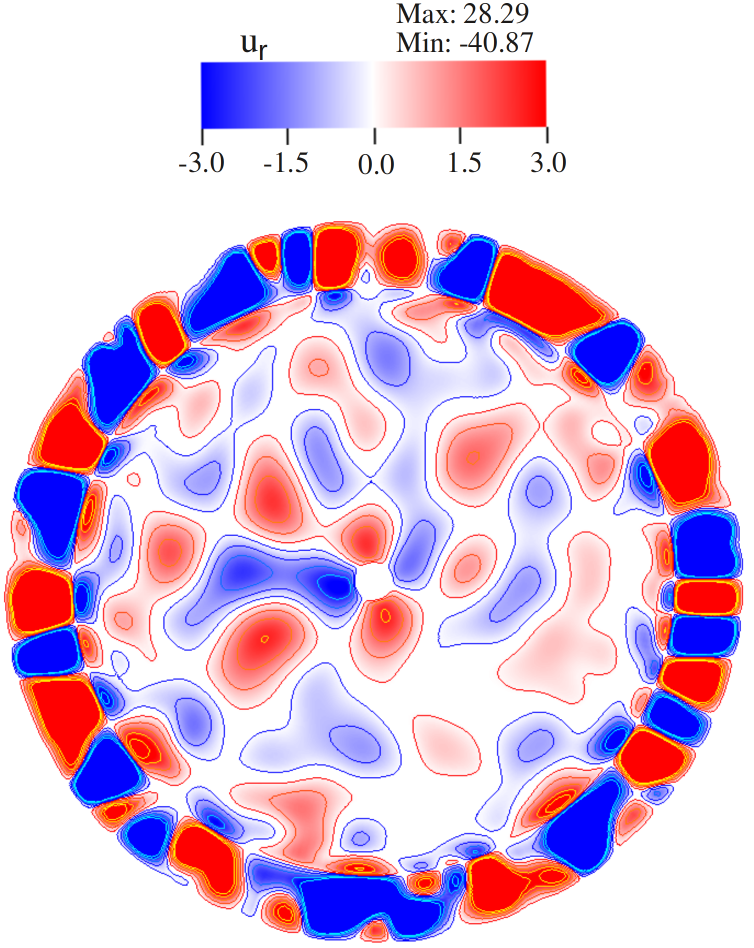} }
\subfloat[]{\includegraphics[angle =0, width=0.44\textwidth]{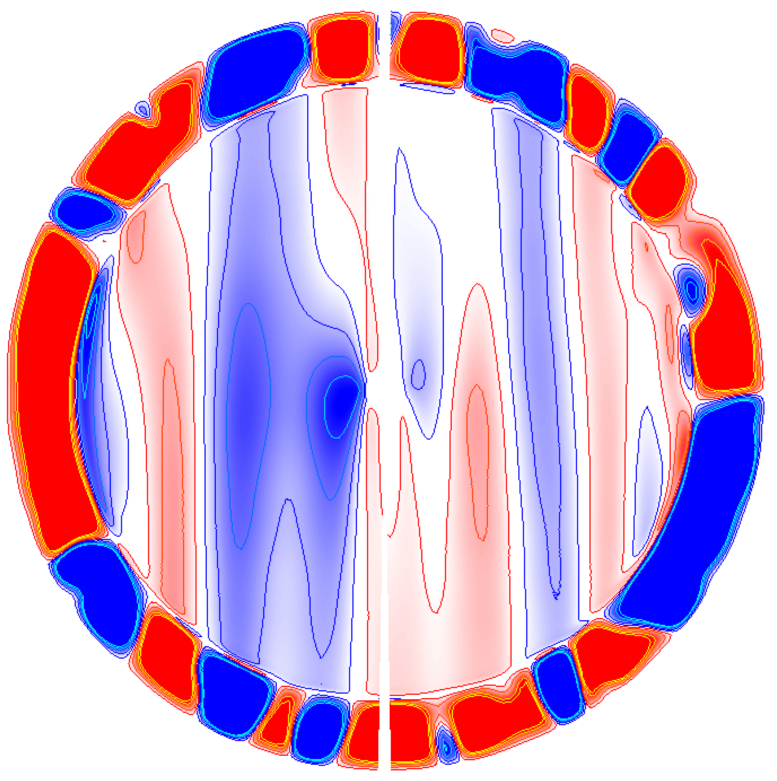} }
\caption{ (a) Equatorial section of $u_r$ and (b) azimuthal section of $u_\phi$
showing a typical
flow pattern for the metallic core-silicate magma ocean system for the default model
$D$.}

\label{fig:fig1}
 \end{center}
\end{figure}

Fig. \ref{fig:fig1} shows equatorial and azimuthal sections for the 
default flow.  For easier comparison between the core and MO, and between
different parameter ratios, we have chosen to fix the max-min
values for the colour scale at $\pm3$ dimensionless units. The actual max-min values are provided above the scale bar.  It is clear that the flow in the MO is much stronger than in the core, and the flow has an obvious cell structure.
From the azimuthal section (Fig. \ref{fig:fig1}b) it is seen
that the core feels much more strongly the effect of rotation than the MO.
Magma ocean convection cells drive extended columnar structures in the core.

\begin{figure} 
 \begin{center}
\subfloat[]{\includegraphics[angle =0, width=0.48\textwidth]{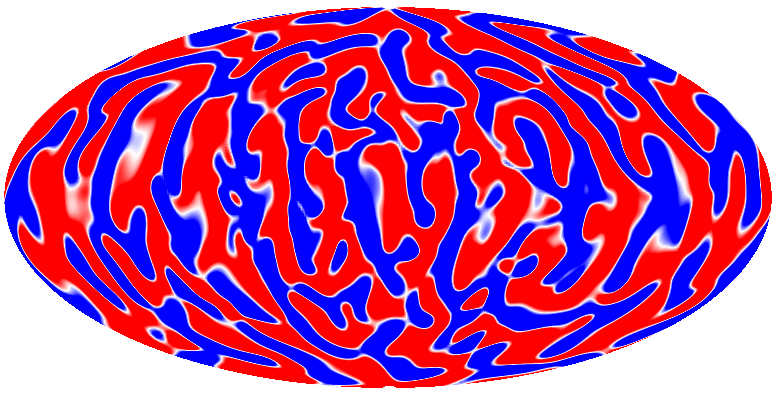} }
\subfloat[]{\includegraphics[angle =0, width=0.48\textwidth]{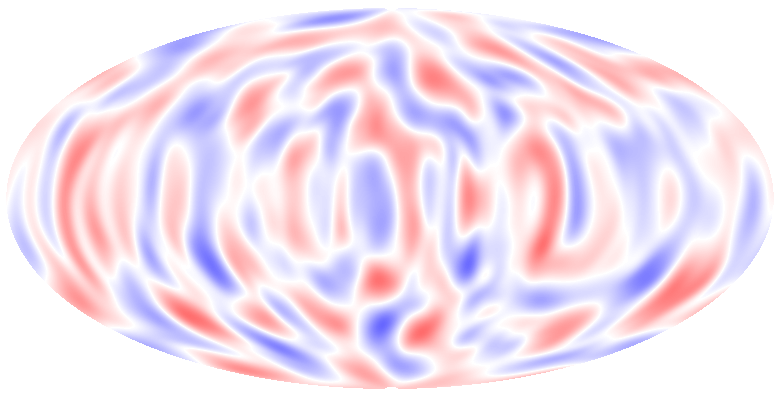} }
\caption{ \small {Radial flow pattern,
$u_r$ at $r=r_l\pm(1/2)(r_o-r_l)$,
inside (a) magma ocean and (b) core
for model $D$.
Same colour scale as Fig.\ \ref{fig:fig1}.
}}
\label{fig:fig2}
 \end{center}
\end{figure}

Radial sections are shown in Fig.\ref{fig:fig2} 
for $r=r_l\pm(1/2)(r_o-r_l)$, i.e.\ in the middle of the MO and the 
same distance below the interface with the core at $r_l$.
The alternating colour cells indicate alternative up- and down-welling in narrow rolls that are similar in both layers -- the core flow appears to imitate the patterns in the MO, but the magnitude of the velocity is much lower. 
We have calculated the \textit{rms} velocities in core and MO separately, denoted $u_c^{RMS}$ and $u_o^{RMS}$ respectively. Their values for this default model are 4.99 and 16. The ratio of the two velocities 
is denoted ${U_{o/c}= u_o^{RMS}/u_c^{RMS}}$.

\subsection{Effect of change in Rayleigh number ($Ra$):}

\begin{figure} [!h]
 \begin{center}
\subfloat[]{\includegraphics[angle =0, width=0.3\textwidth]{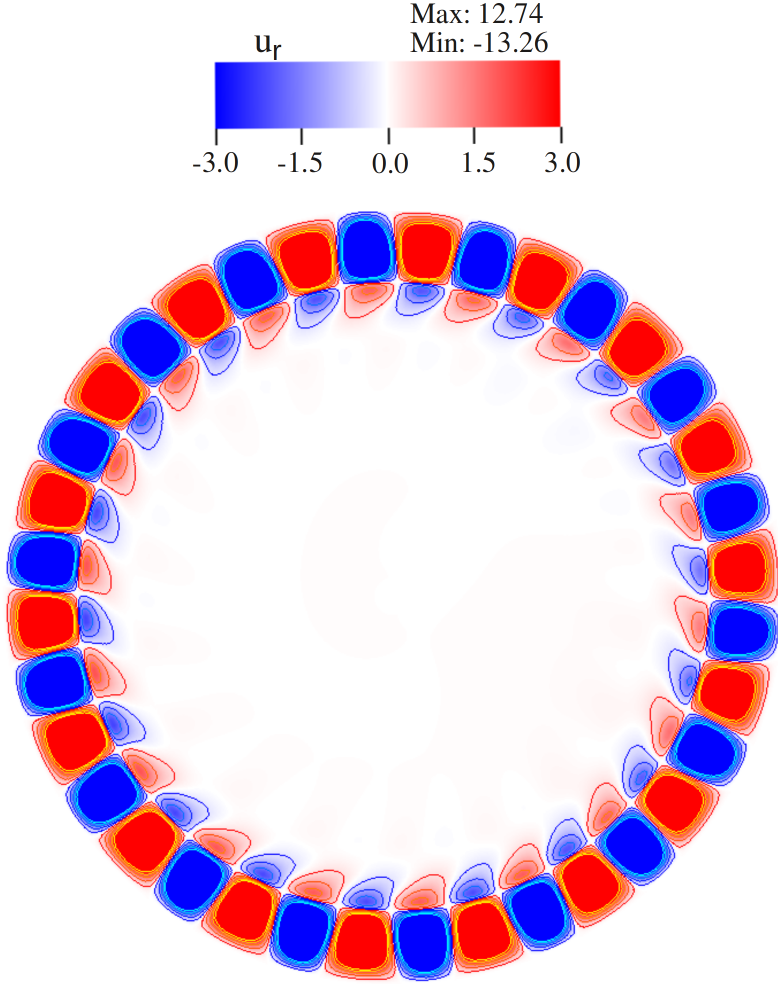} }
\subfloat[]{\includegraphics[angle =0, width=0.3\textwidth]{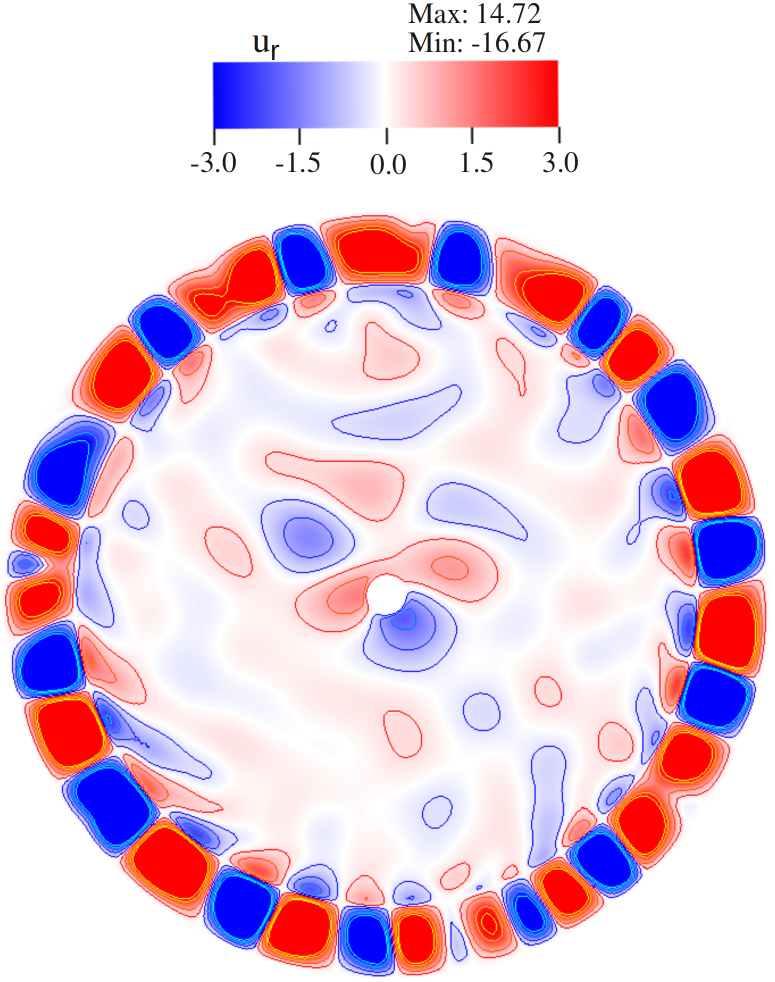} }
\subfloat[]{\includegraphics[angle =0, width=0.31\textwidth]{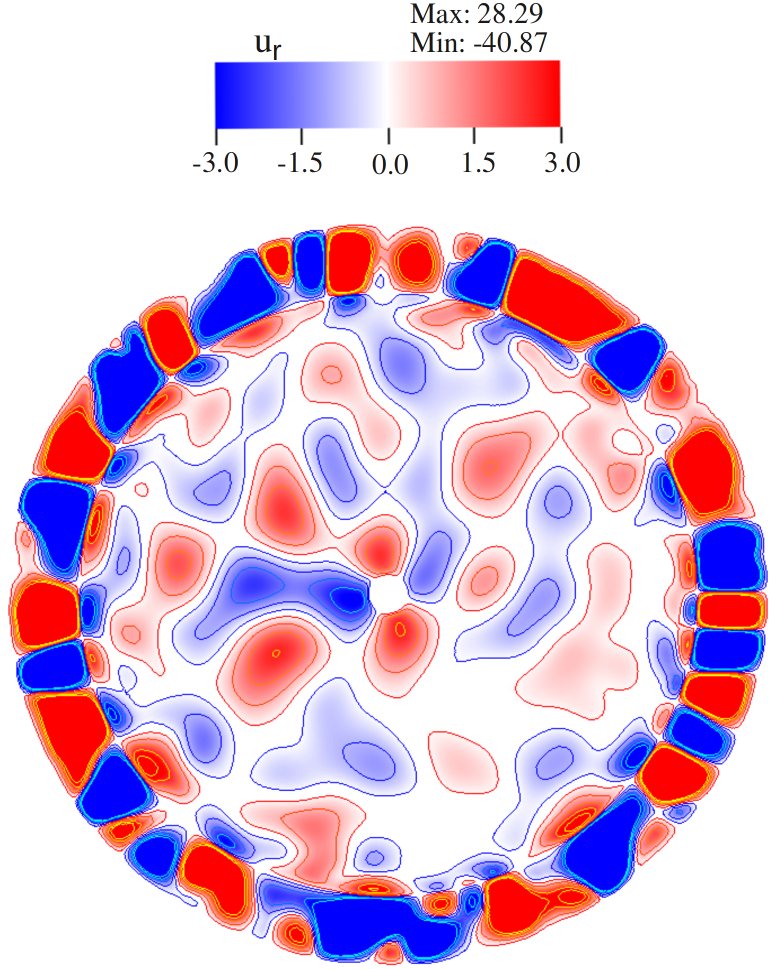} }

\subfloat[]{\includegraphics[angle =0, width=0.3\textwidth]{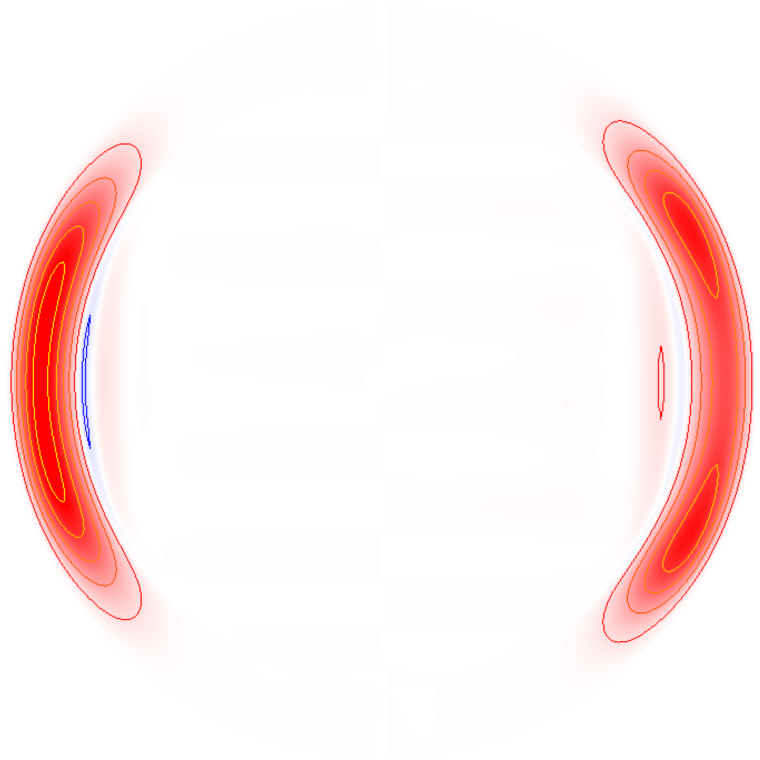} }
\subfloat[]{\includegraphics[angle =0, width=0.3\textwidth]{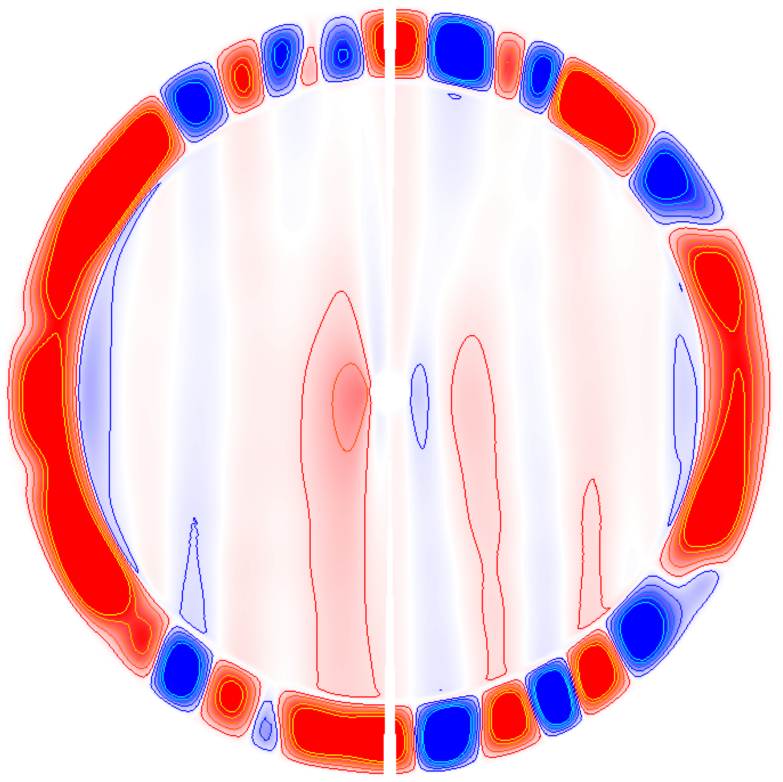} }
\subfloat[]{\includegraphics[angle =0, width=0.3\textwidth]{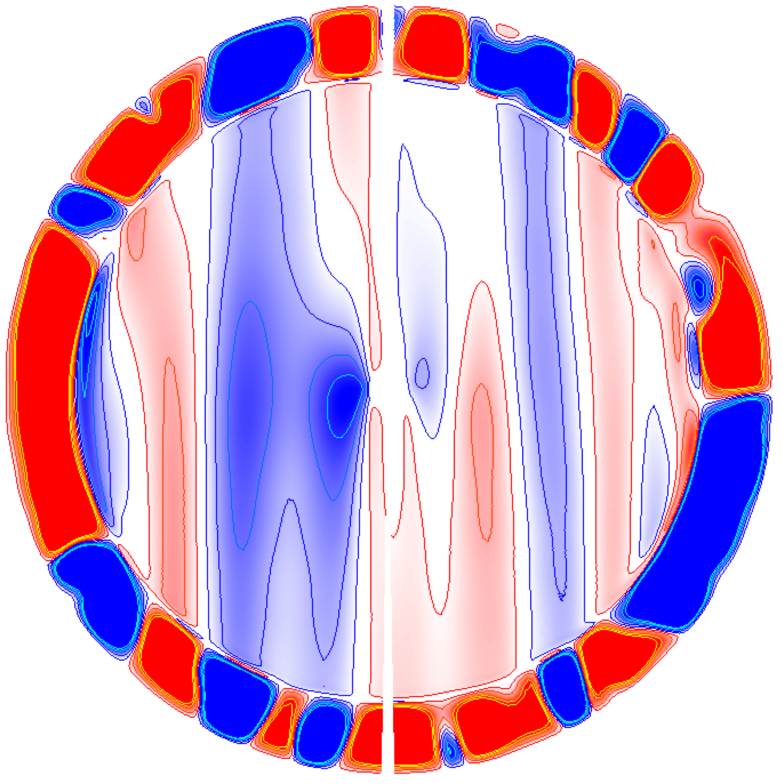} }
\caption{
Equatorial (upper) and azimuthal (lower) sections for  (a) $Ra\_20$, (b) $Ra\_50$, and (c) 
$Ra\_200$ $\equiv$ $D$.
The figures represent $u_r$ and $u_\phi$ for equatorial and azimuthal sections, respectively.}
\label{fig:fig3}
\end{center}
\end{figure}

We have performed simulations to study the interaction between the layers with changing Rayleigh number. The flow patterns are presented in Fig.\ref{fig:fig3} for different super-critical $Ra$ from $20$ to $200$, where
$Ra_c\approx4.7$.

At $Ra$ almost 5 times critical ($Ra=20$) the nonlinear 
flow is still very regular.
In Fig.\ref{fig:fig3}a, we find that 
there is no flow in the 
core other than that near the interface,
where flow gathering at the base of an upwelling in the MO drives a weak downwelling into the core below.
At larger $Ra$, there is stronger downward 
penetration, parallel to the axis, of the MO flow into the core, seen in the azimuthal sections.
Other than this, increase in $Ra$ has a typical effect on both the layers -- an increase in chaotic motion. At higher $Ra$,
convection in the MO is more vigorous,
breaking the
symmetry in the shape and size of the MO cells. Secondary cells are formed, seen in Fig.\ref{fig:fig3}(c) $Ra=200$, 
but coupling remains the same, with 
rotationally-constrained continuation of $u_\phi$ into the core, 
driving a stronger core flow.
As $Ra$ increases from 20 to 200, $u_o^{RMS}$ increases substantially, but $U_{o/c}$ decreases from 5.83 to 3.2.
In the following, we vary other parameters to examine their effect on
the core-ocean coupling.  As they affect the supercriticality of the flow,
this largest value for $Ra$ is taken for the default parameter set.

\begin{figure} [!h]
 \begin{center}
\subfloat[]{\includegraphics[angle =0, width=0.3\textwidth]{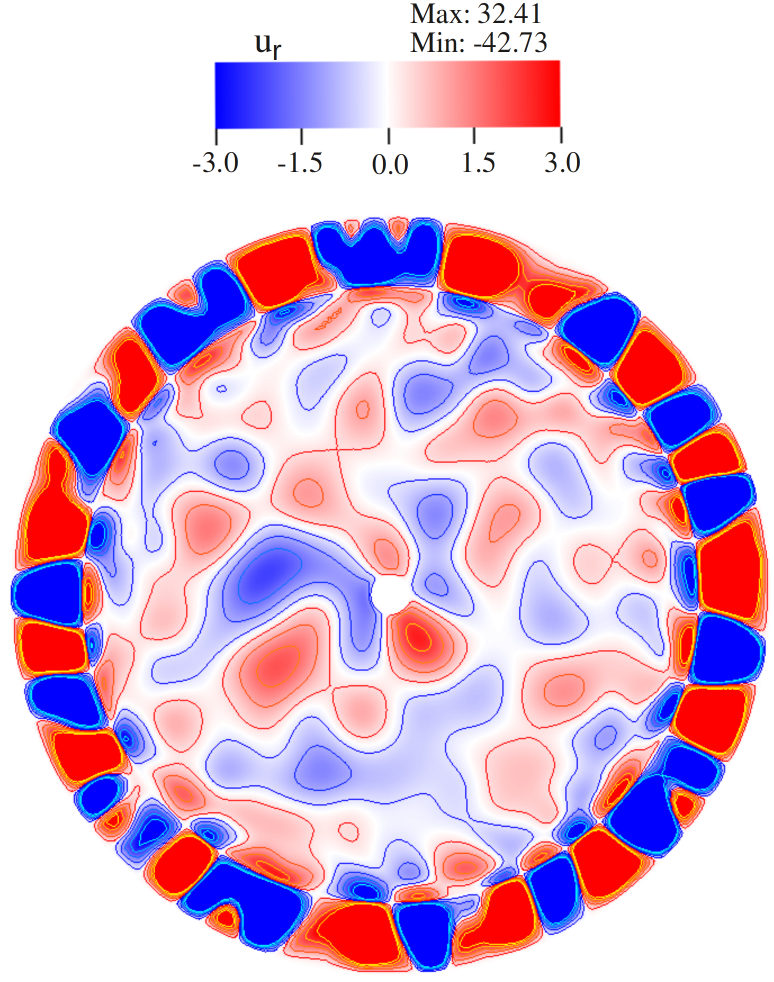} }
\subfloat[]{\includegraphics[angle =0, width=0.31\textwidth]{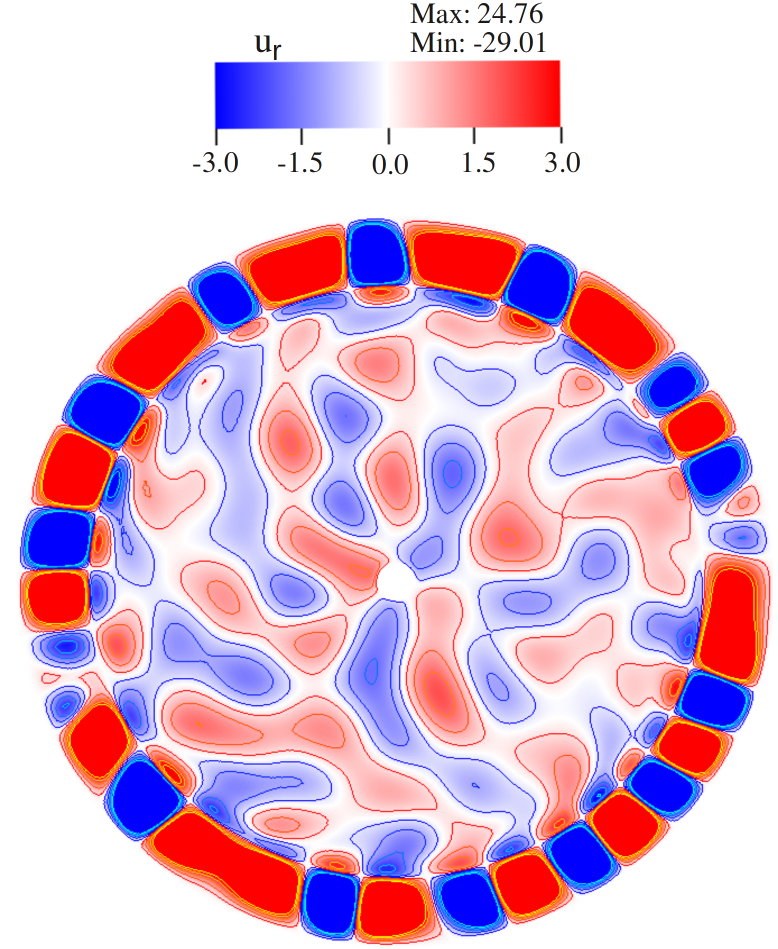} }
\subfloat[]{\includegraphics[angle =0, width=0.3\textwidth]{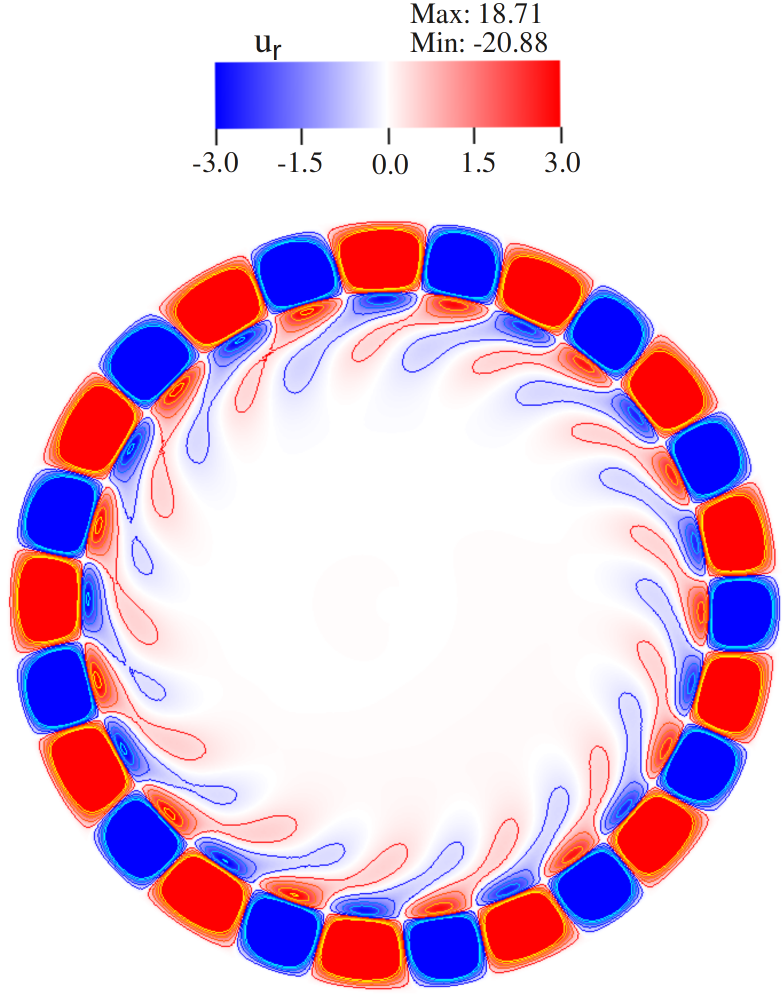} }

\subfloat[]{\includegraphics[angle =0, width=0.31\textwidth]{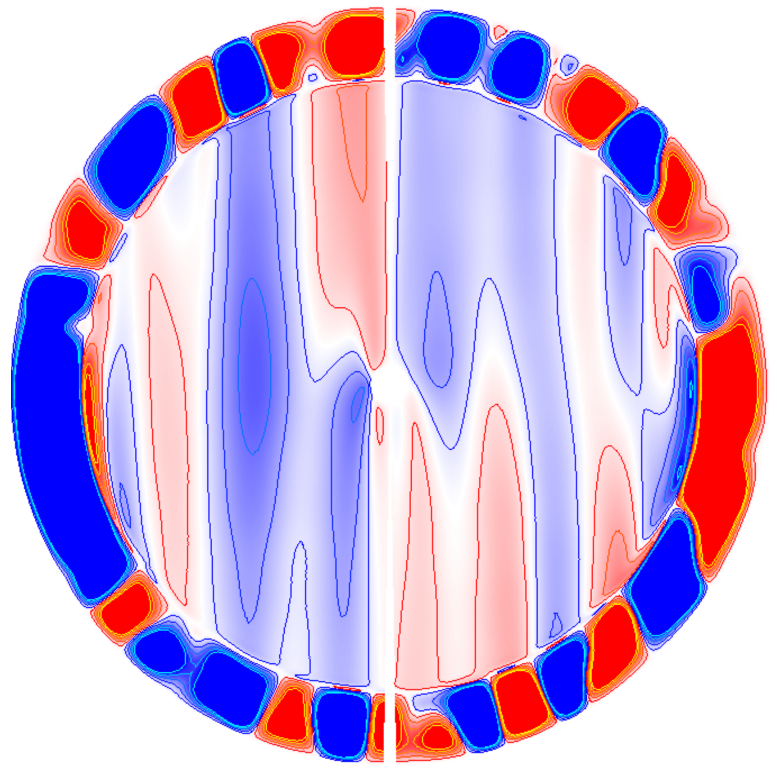} }
\subfloat[]{\includegraphics[angle =0, width=0.31\textwidth]{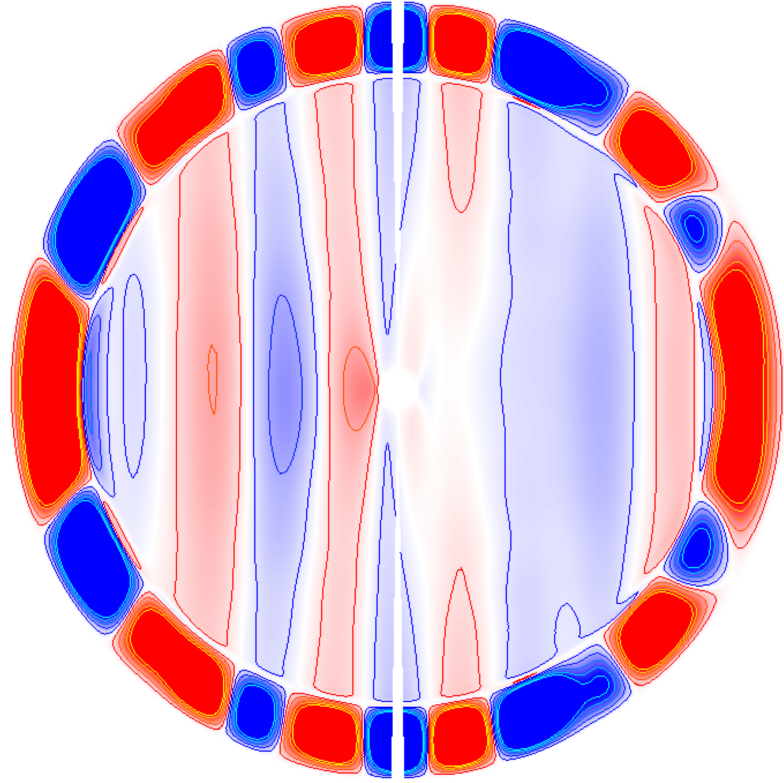} }
\subfloat[]{\includegraphics[angle =0, width=0.31\textwidth]{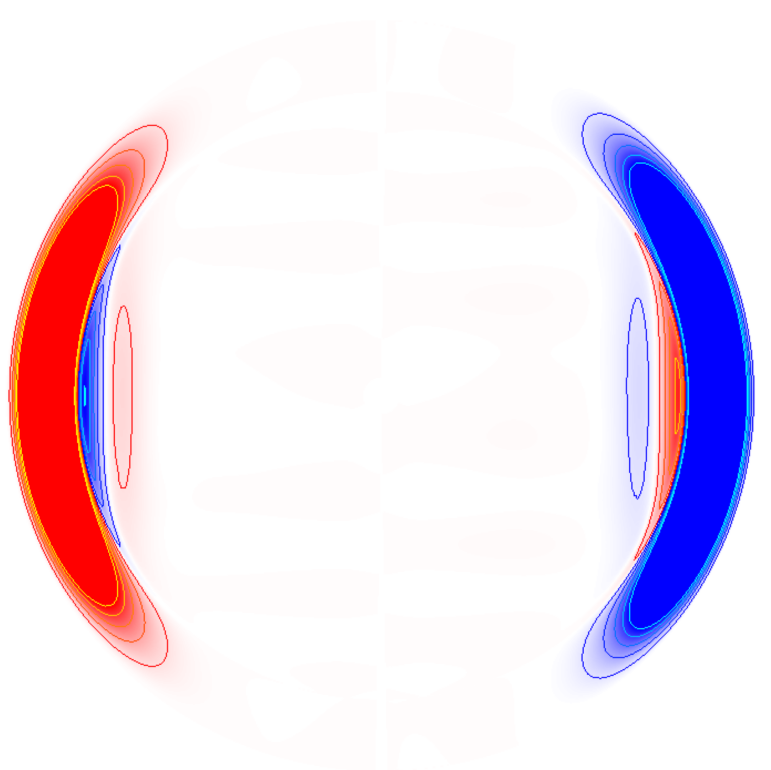} }
\caption{ Change in equatorial and azimuthal flow pattern with increasing thermal diffusivity ratio (a) 
$D~(\kappa_{o/c}=0.1)$
(b) $\kappa\_0.3$, (c) $\kappa\_0.5$. The figures show $u_r$ and $u_\phi$ for equatorial and azimuthal sections, respectively.
}
\label{fig:fig4}
\end{center}
\end{figure}

\subsection{Effect of increasing thermal diffusivity ratio ($\kappa_{o/c}$):}

Starting from the standard model with $\kappa_{o/c}=0.1$, we increased the ratio in small increments. We found that the $\kappa$ ratio between the layers has a huge effect on the flows of both the layers. In the MO, a lower ratio ($\kappa_{o/c}=0.1$) (Fig.\ref{fig:fig4}a) caused a chaotic convection pattern with unequal convection cells. Some of the cells are even broken into smaller interactive cells. Increasing the value of $\kappa_{o/c}$ increases the of $\kappa$ in the MO, which increases the  contribution of conduction in maintaining the heat flux, hence decreasing convective vigour (Fig.\ref{fig:fig4}). The increased $\kappa_{o/c}$ reduces temperature gradients in the MO, and therefore reduces $u_r$. 
Differences in the number of cell might be attributed to a different initial condition.
At $\kappa_{o/c}= 0.5$, the MO cells are regular and represent a steady convective flow (Fig.\ref{fig:fig4}c), and the core flow is very weak, with perturbations only due to the MO flow.

\begin{figure} 
 \begin{center}

\subfloat[]{\includegraphics[angle =0, width=0.32\textwidth]{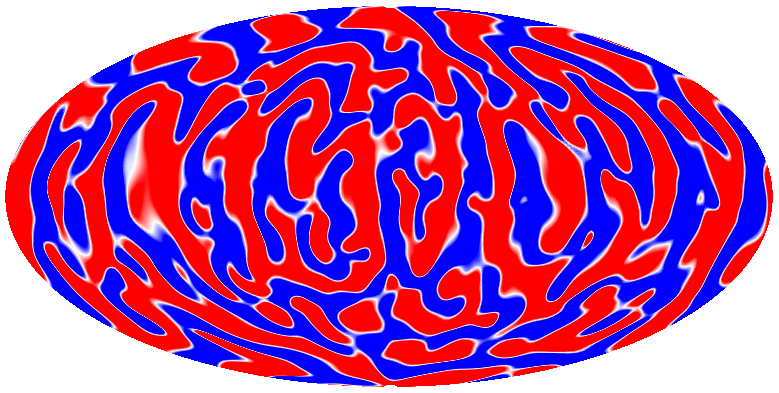} }
\subfloat[]{\includegraphics[angle =0, width=0.32\textwidth]{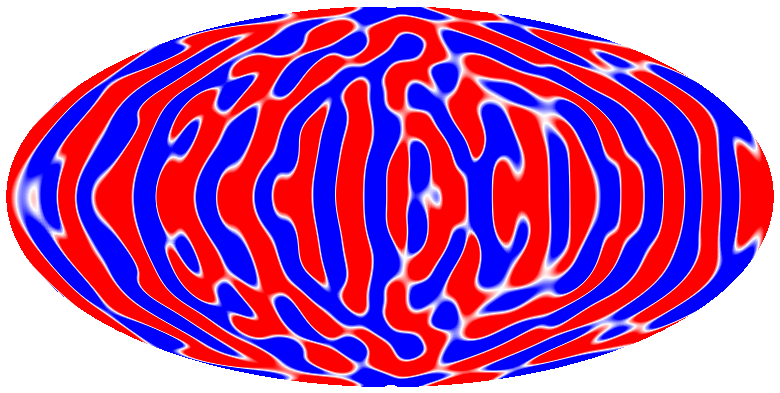} }
\subfloat[]{\includegraphics[angle =0, width=0.32\textwidth]{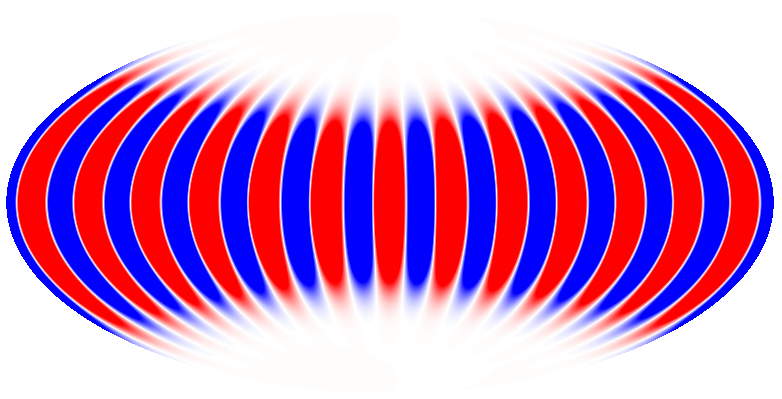} }
\caption {Band formation in radial section at $r= r_o + 1/2(r_o-r_l$) with changing thermal diffusivity ratio (a) $\kappa\_0.3$, (b) $\kappa\_0.4$, (c) $\kappa\_0.5$. Colours show $u_r$ with same colour scale as Fig.\ \ref{fig:fig4}.}
\label{fig:fig5}
 \end{center}
\end{figure}

The corresponding azimuthal sections (Fig.\ref{fig:fig4}d,e,f) also show the change in relative energy of the MO flow versus the core as $\kappa_{o/c}$ is increased. 
Organized convection in the MO results in the formation of strong, disconnected columns, seen in the radial 
\begin{figure} [!htb]
 \begin{center}

\subfloat[]{\includegraphics[angle=0, width=0.5\textwidth]{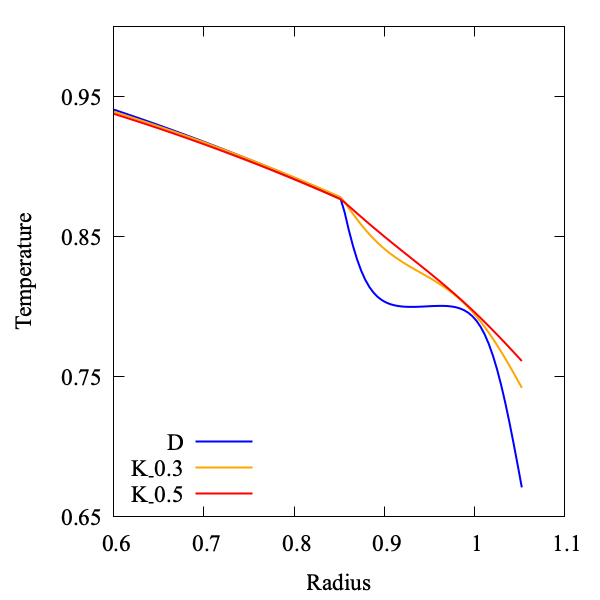} }
\caption{ \small {Snapshot of radial temperature distribution for varying $\kappa_{o/c}$.  Model in key, 
where $\kappa_{o/c}=0.1$ for $D$.
}}
\label{fig:fig6}
 \end{center}
\end{figure}
section (Fig.\ref{fig:fig5}c) unlike those connected weak columns at lower $\kappa_{o/c}$ values (Fig.\ref{fig:fig5}a).  The transition is visible in Fig.\ref{fig:fig5}b where few bands have formed. The velocity drops significantly
towards the poles (Fig.\ref{fig:fig5}c).
The range for the ratio $U_{o/c}$ 
increases from 3.2 to 4.98
as $\kappa_{o/c}$ is increased, 
due to weakening of the core flow. The chaotic convective flows in the MO at lower $\kappa$ ratio helps thermal distribution within the layer, seen in
Fig.\ref{fig:fig6},
and the MO is expected to cool faster at $\kappa_{o/c}=0.1$ than at $\kappa_{o/c}=0.5$.  

Interesting points can be noted when the change in $\kappa_{o/c}$ is compared to the cases where $Ra$ was changed. Even though $Ra$ is fixed at 200 (the value for the default model $D$), for all the cases of $\kappa$ studied, the flow patterns and flow interactions for increasing $\kappa_{o/c}$ are quite similar to the patterns found when $Ra$ is decreased.  Fig.\ref{fig:fig4}c is very similar to that found in Fig.\ref{fig:fig3}a. The convection cells in the MO are very regularly distributed in both cases, but 
the numbers of cells are significantly different 
- Fig.\ref{fig:fig3}a has 36 cells and Fig.\ref{fig:fig4}c has 26. 
\begin{figure} [!h]
 \begin{center}
\subfloat[]{\includegraphics[angle =0, width=0.3\textwidth]{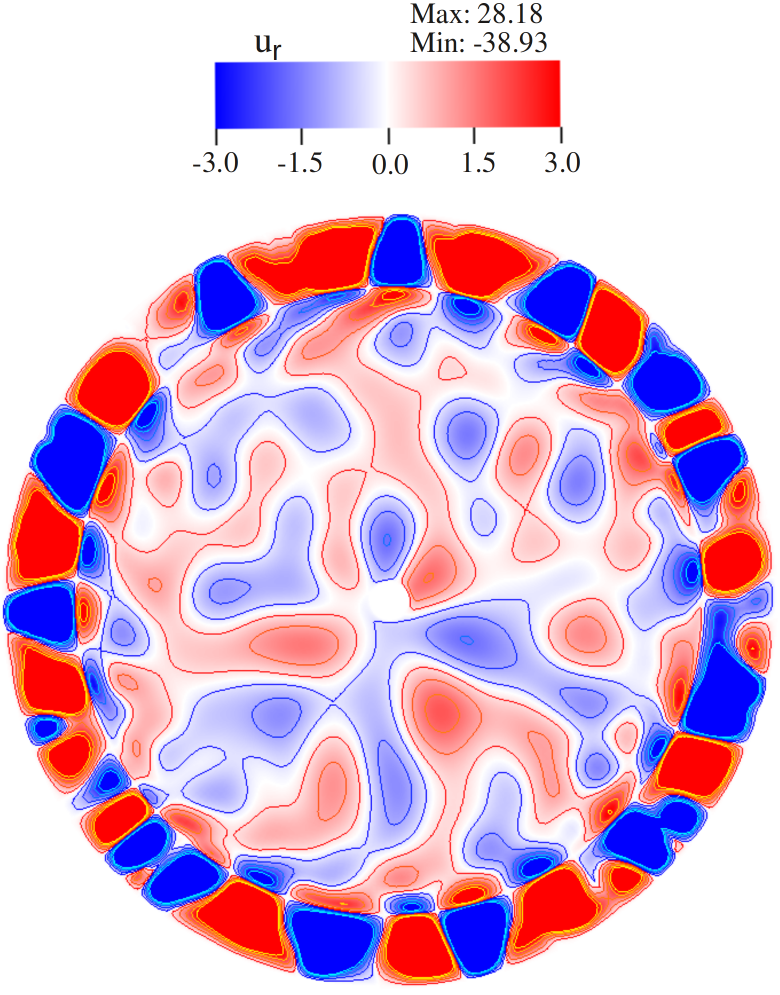} }
\subfloat[]{\includegraphics[angle =0, width=0.3\textwidth]{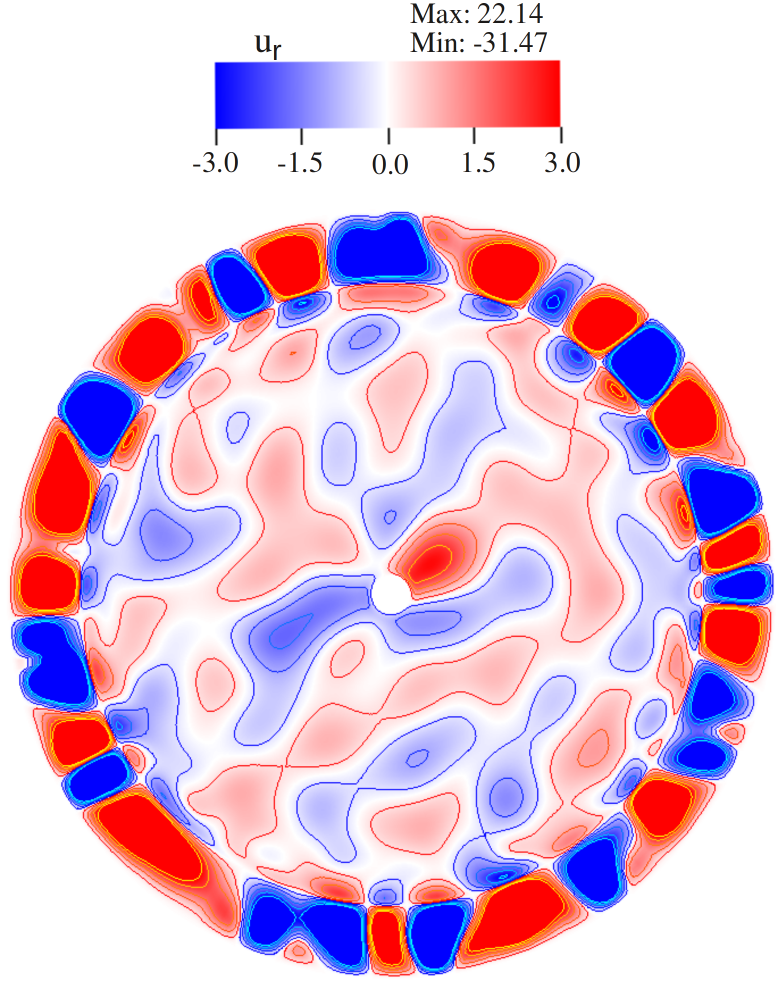} }
\subfloat[]{\includegraphics[angle =0, width=0.3\textwidth]{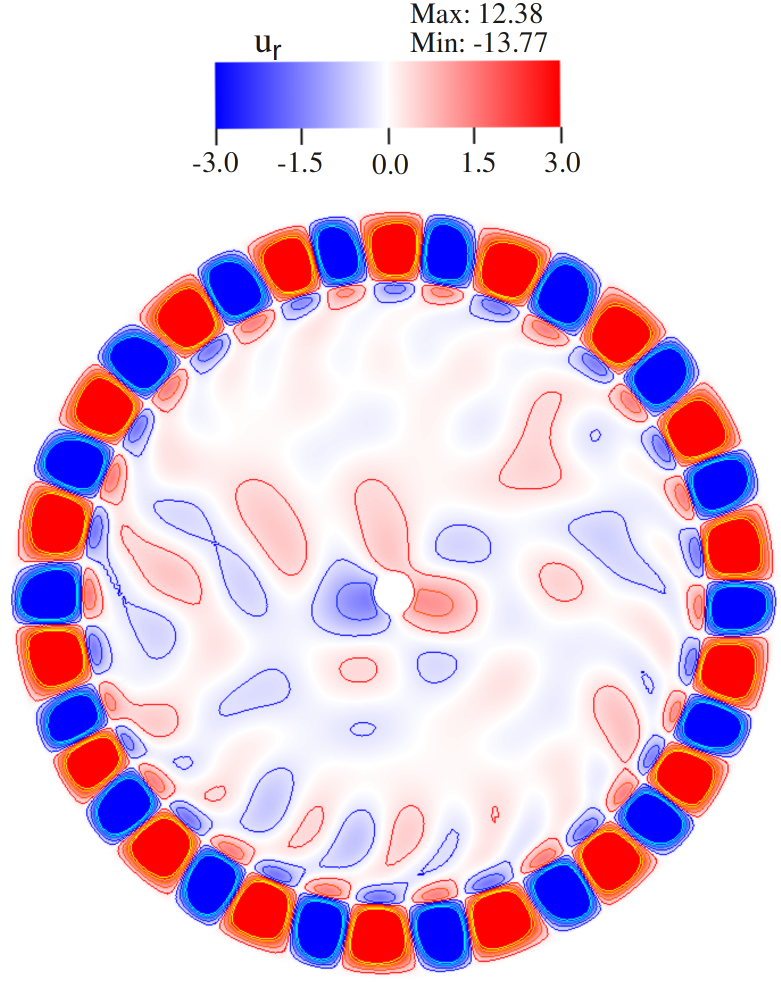} }

\subfloat[]{\includegraphics[angle =0, width=0.3\textwidth]{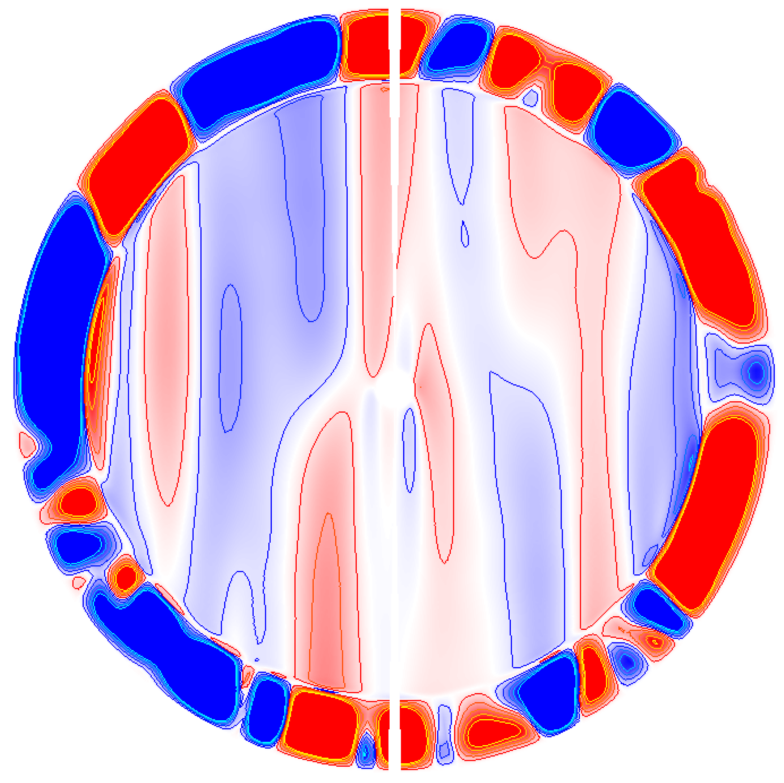} }
\subfloat[]{\includegraphics[angle =0, width=0.3\textwidth]{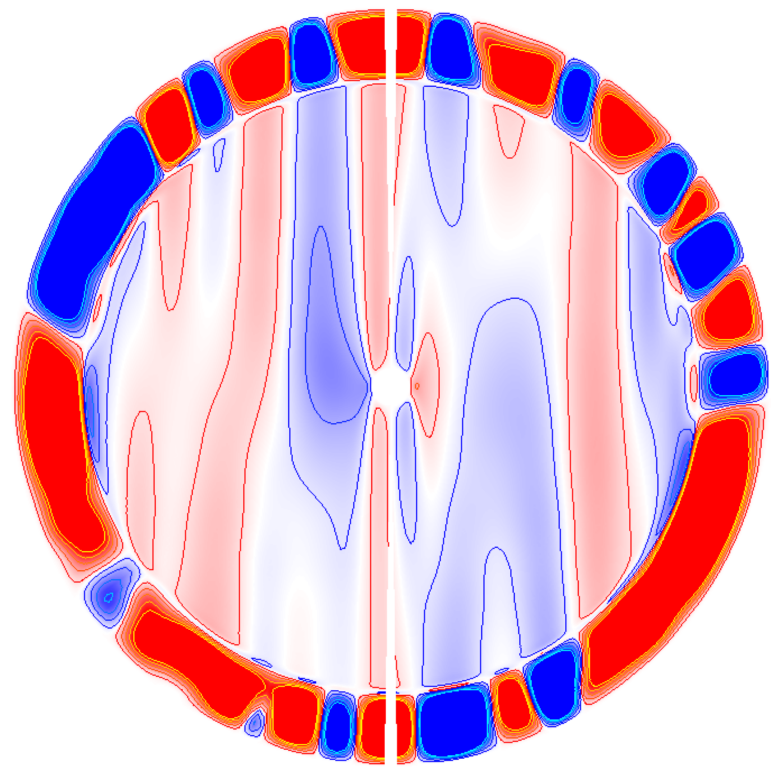} }
\subfloat[]{\includegraphics[angle =0, width=0.3\textwidth]{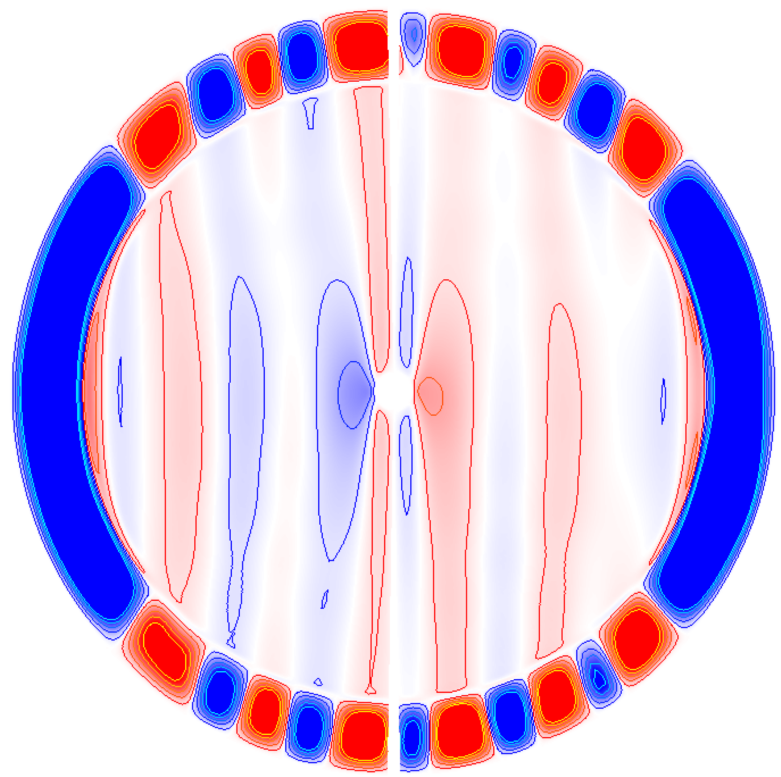} }
\caption{ Effect of decreasing thermal expansion coefficient ratio.
Equatorial and azimuthal sections for 
(a) $D~(\alpha_{o/c}=5)$, 
(b) $\alpha\_3$, (c) $\alpha\_1$. 
The figures show $u_r$ and $u_\phi$ for equatorial and azimuthal sections respectively.
}

\label{fig:fig7}
\end{center}
\end{figure}
Given the fact that Fig.\ref{fig:fig4}c has higher $Ra$, it is uncertain what might cause the formation of more convection cells for the case of Fig.\ref{fig:fig3}a. The radial sections for both cases are also almost identical except the number of columns.

\subsection{Effect of decreasing thermal expansion coefficient ratio ($\alpha_{o/c}$):}

Starting from  $\alpha_{o/c}=5$, we ran a number of simulations until both the layers have equal thermal expansion coefficient (i.e. $\alpha_{o/c} =1$). Diminishing the difference between $\alpha$ of the two layers have similar qualitative effect 
(Fig.\ref{fig:fig7}) on their relative flow energy for increasing the 
$\kappa_{o/c}$.  It should be noted from equation (\ref{eq:mom}) that a change in $\alpha_{o/c}$ is directly equivalent to changing $Ra$ for the MO.

Therefore, as the ratio is brought down to $\alpha_{o/c}=1$,
both the MO flow and 
the core flow weaken 
(Fig.\ref{fig:fig7}c). Irregular shaped MO convection cells also transformed into regular closely compact cells. Nevertheless, these cells are more elongated and as a result we have higher number of cells compactly placed within the same thickness of the magma ocean. Azimuthal sections (Fig.\ref{fig:fig7}d,e,f) show strong effect of rotation on core flow for cases $\alpha_{o/c}$ = 3 and 1, with corresponding flow just above and beneath the interface. 

\begin{figure} [!htb]
 \begin{center}
 
\subfloat[]{\includegraphics[angle =0, width=0.32\textwidth]{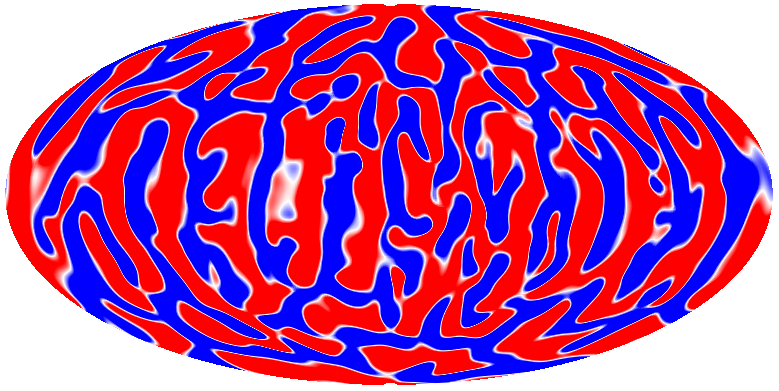} }
\subfloat[]{\includegraphics[angle =0, width=0.32\textwidth]{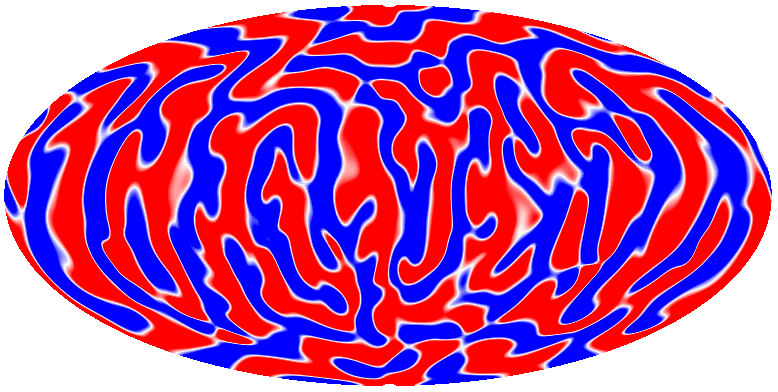} }
\subfloat[]{\includegraphics[angle =0, width=0.32\textwidth]{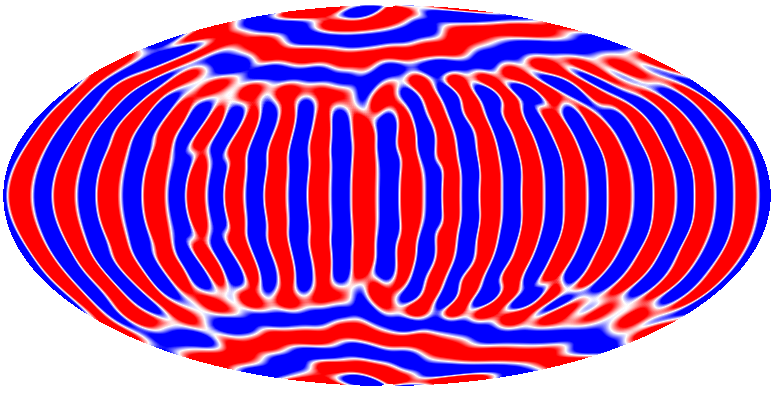} }

\caption { Effect of decreasing thermal expansion coefficient ratio.
Radial sections taken at $r= r_o + (1/2)(r_o-r_l$). 
(a) $D$, (b) $\alpha\_3$, (c) $\alpha\_1$. Colours show $u_r$ with same colour scale as Fig.\ \ref{fig:fig7}.}

\label{fig:fig8}
 \end{center}
\end{figure}

In radial sections (Fig.\ref{fig:fig8}) we find formation of similar strong columns with decreasing $\alpha_{o/c}$ as found for increasing $\kappa_{o/c}$ (Fig.\ref{fig:fig5}).  But unlike Figs.\ref{fig:fig5}b \& c, here we observe strong flows around the polar region, and interestingly, the columns are terminated by polar bands (Fig.\ref{fig:fig8}c). 
This type of multiple polar band is not very common in rotating spherical shell convection. In their recent work on polar waves in such systems, \cite{garcia2019polar} encountered polar bands, and in special cases a combination of band and equatorial convection. They attributed this appearance to a change in Floquet multipliers (FM) of equatorially symmetric (ES) rotating waves. This might also be comparable with the stratified layers found inside the core in some studies \citep{mound2019regional, hardy2020enhanced}.

\begin{figure} [!htb]
 \begin{center}
{\includegraphics[angle =0, width=0.5\textwidth]{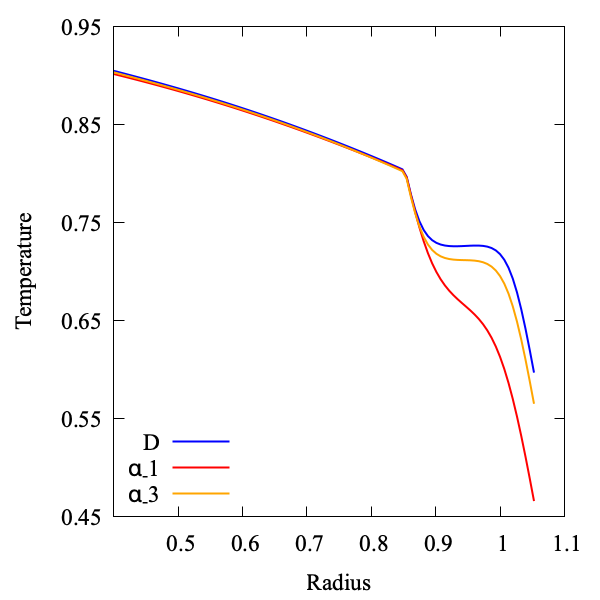} }
\caption{ \small { Snapshot of radial temperature distribution for varying 
$\alpha_{o/c}$.  Model in key, where $\alpha_{o/c}=5$ for $D$.
}}

\label{fig:fig9}
 \end{center}
\end{figure}

Decrease in  $\alpha_{o/c}$ leads to further suppression of core flow relative to the MO flow, where $U_{o/c}$ value increases from 3.2 to 3.68 to 4.19 as $\alpha_{o/c}$ is reduced from 5 to 3 to 1.
Temperature profiles
are plotted in 
Fig.\ref{fig:fig9}.
They are essentially invariant with respect to $\alpha_{o/c}$ within the core, as the core flow is so weak.
It shows that the magma ocean cooling follows a consistent form while the ratio drops to $\alpha_{o/c}$=3 from =5, but ends at slightly lower temperature.  But the slope 
is steeper at $\alpha_{o/c}=1$ making the top of the MO much cooler. 
Increase in $\alpha_{o/c}$ may be inferred as higher value of $\alpha$ in the MO i.e. larger $Ra$. This should result in vigorous convection in the MO which can carry the heat out more easily and eventually causing larger temperature drop across the core-MO boundary. We can see this pattern in temperature distribution at fixed $Ra$ (=200) and $\delta_{o/c}$ (= 0.25)(Fig.\ref{fig:fig9}) where highest drop in temperature (from 0.83-0.79) is identified for $\alpha_{o/c}=5$.

\subsection{Effect of decreasing viscosity ratio ($\nu_{o/c}$):}

\begin{figure} [!h]
 \begin{center}
\subfloat[]{\includegraphics[angle =0, width=0.35\textwidth]{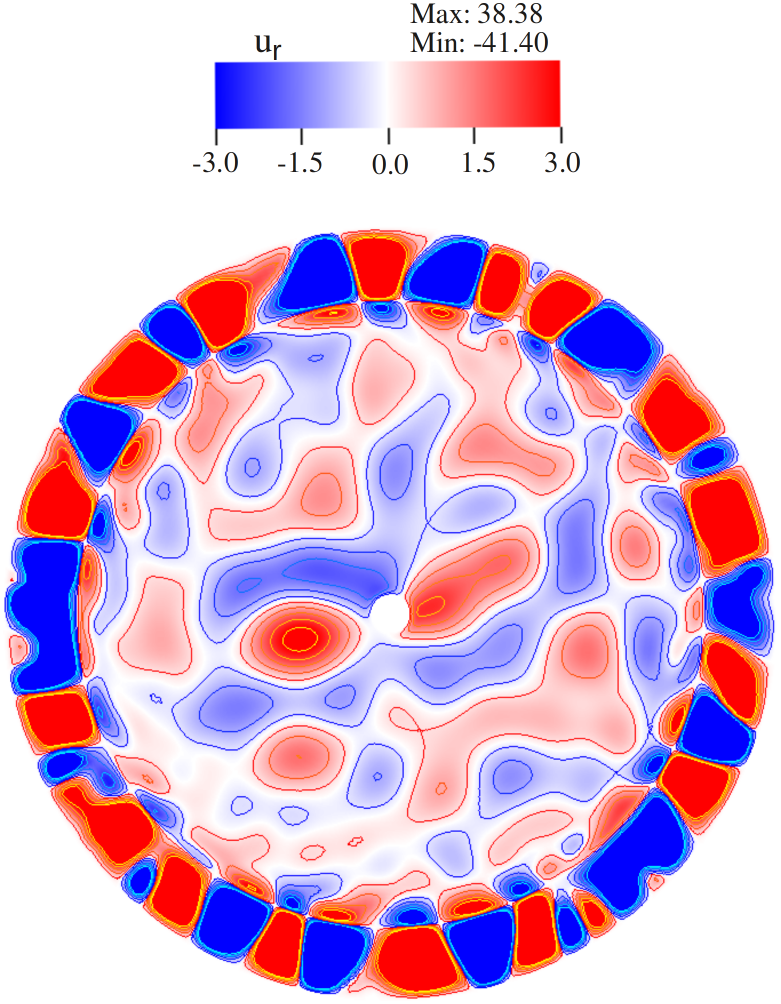} }
\subfloat[]{\includegraphics[angle =0, width=0.35\textwidth]{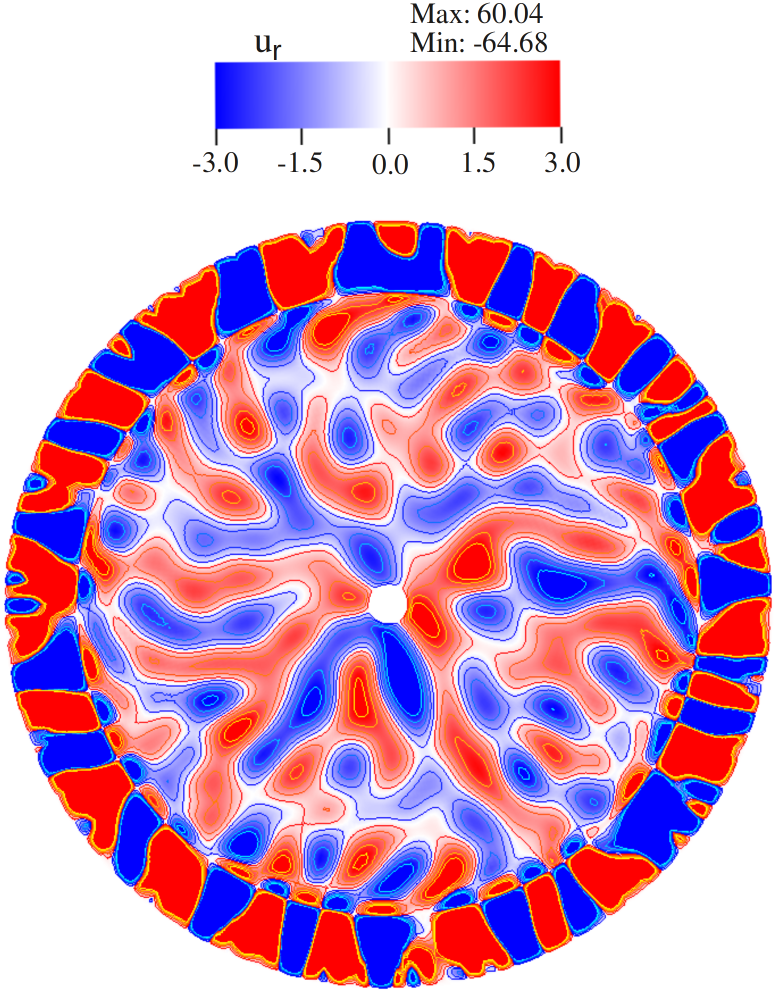} }

\subfloat[]{\includegraphics[angle =0, width=0.35\textwidth]{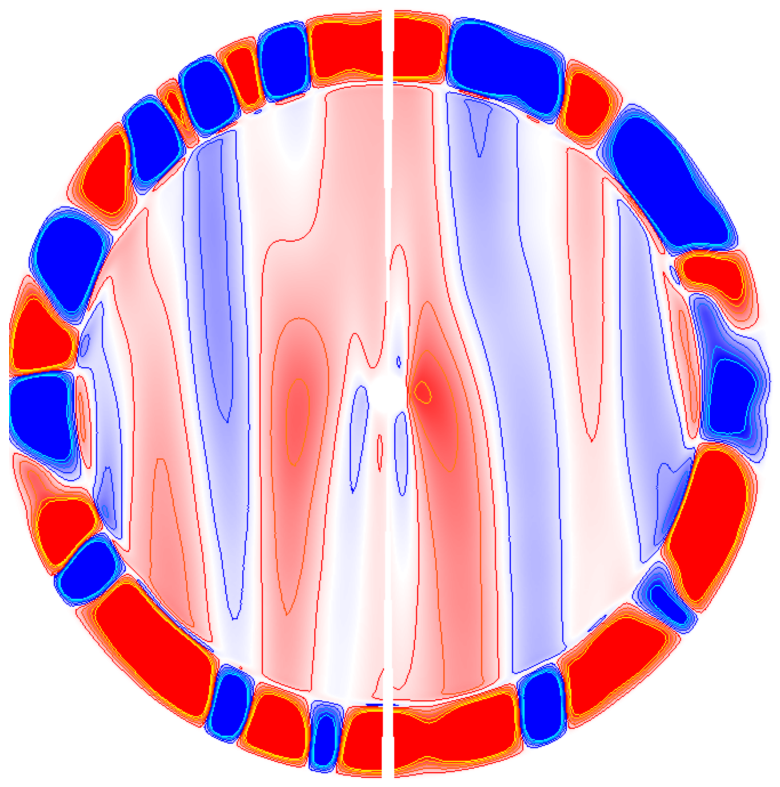} }
\subfloat[]{\includegraphics[angle =0, width=0.35\textwidth]{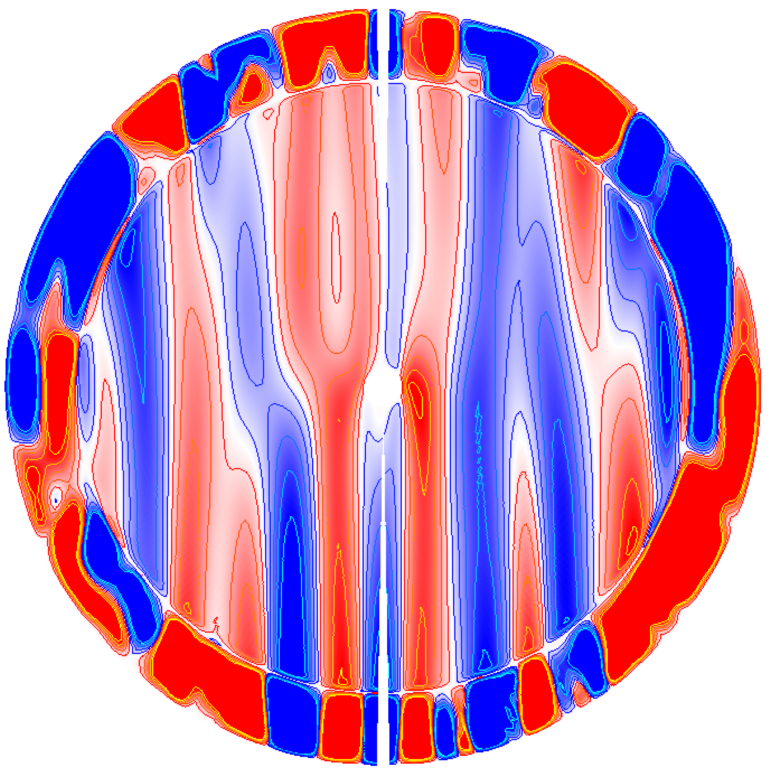} }

\caption{ Equatorial sections (a),(b) of $u_r$ and azimuthal sections (c),(d) of $u_\phi$.  
(a),(c): 
$D~(\nu_{o/c}=25)$,
(b),(d): $\nu\_5$.
}

\label{fig:fig10}
\end{center}
\end{figure}

\begin{figure} [!h]
 \begin{center}
\subfloat[]{\includegraphics[angle =0, width=0.4\textwidth]{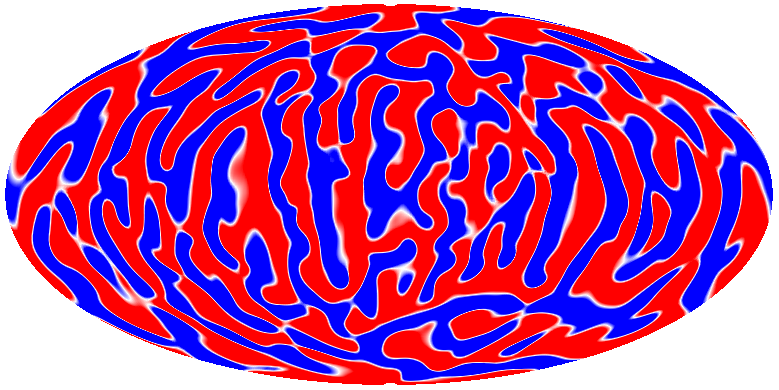} }
\subfloat[]{\includegraphics[angle =0, width=0.4\textwidth]{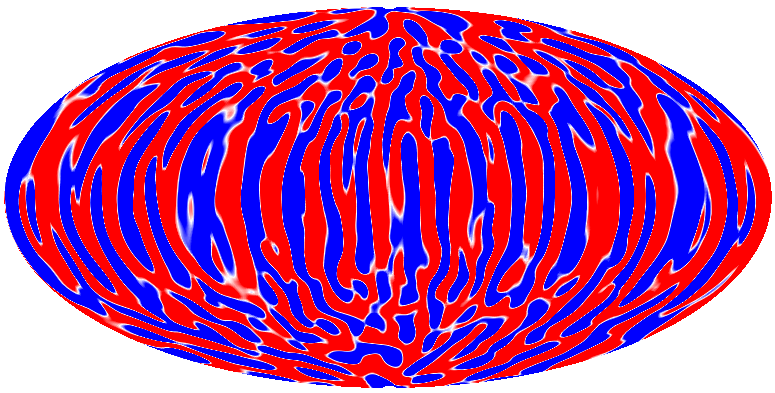} }
\caption{ \small {Radial sections at $r= r_l + 1/2(r_o-r_l$). Left: 
$D~(\nu_{o/c}=25)$,
Right: $\nu\_5$.  Colours show $u_r$ with same colour scale as Fig.\ \ref{fig:fig10}.}}
\label{fig:fig11}
 \end{center}
\end{figure}
 The viscosity is a very important parameter guiding the flow regime of any particular fluid layer. We considered decreasing 
$\nu_{o/c}$ to see 
how the ratio affects the coupling between the core and MO. 
Fig.\ref{fig:fig10} includes the equatorial and azimuthal sections of two cases with $\nu_{o/c}$ =25 \& =5. As we can see, the core flow  becomes much stronger in the latter case (Fig.\ref{fig:fig10} b,d). 

\begin{figure} [!htb]
 \begin{center}
\subfloat[]{\includegraphics[angle =0, width=0.3\textwidth]{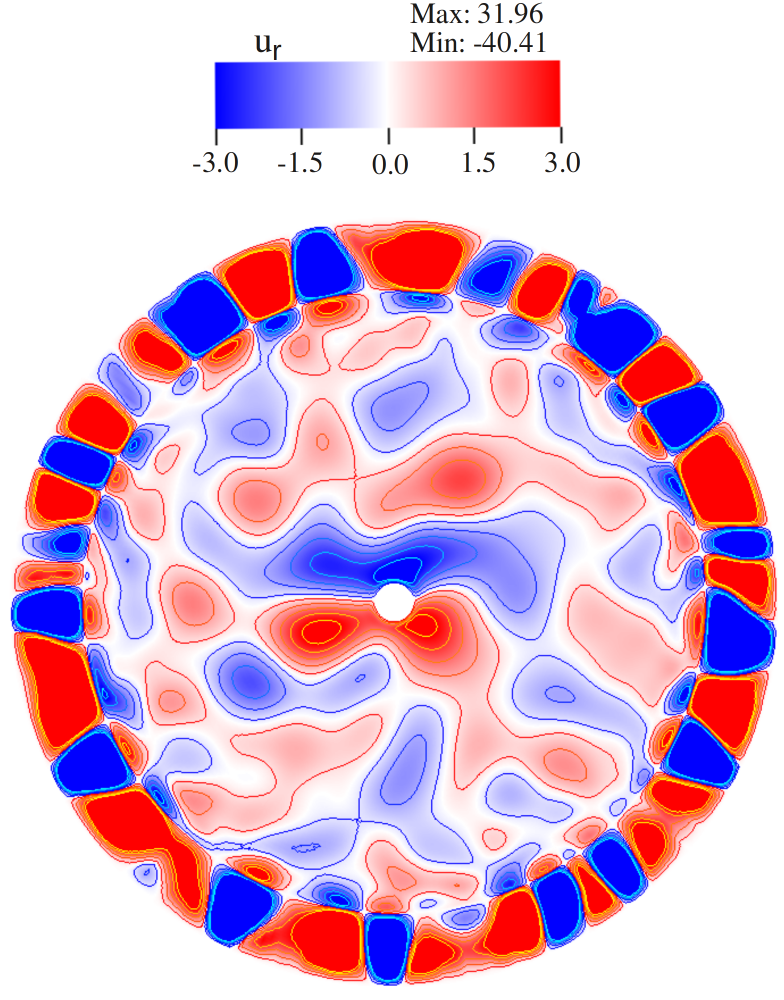} }
\subfloat[]{\includegraphics[angle =0, width=0.3\textwidth]{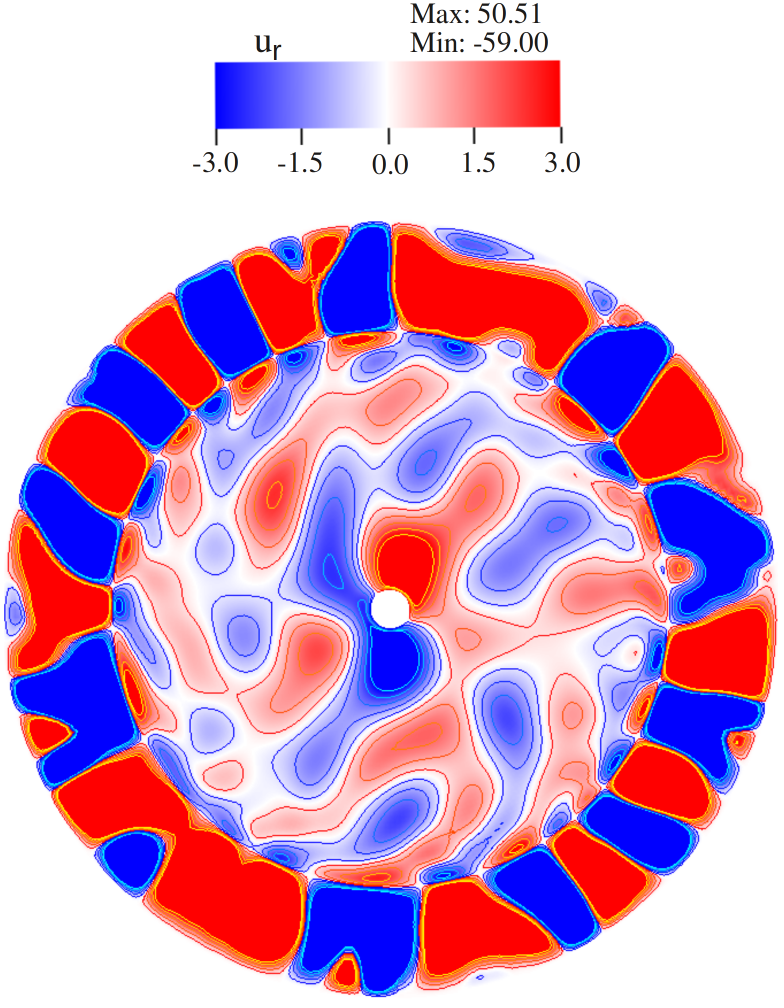} }
\subfloat[]{\includegraphics[angle =0, width=0.3\textwidth]{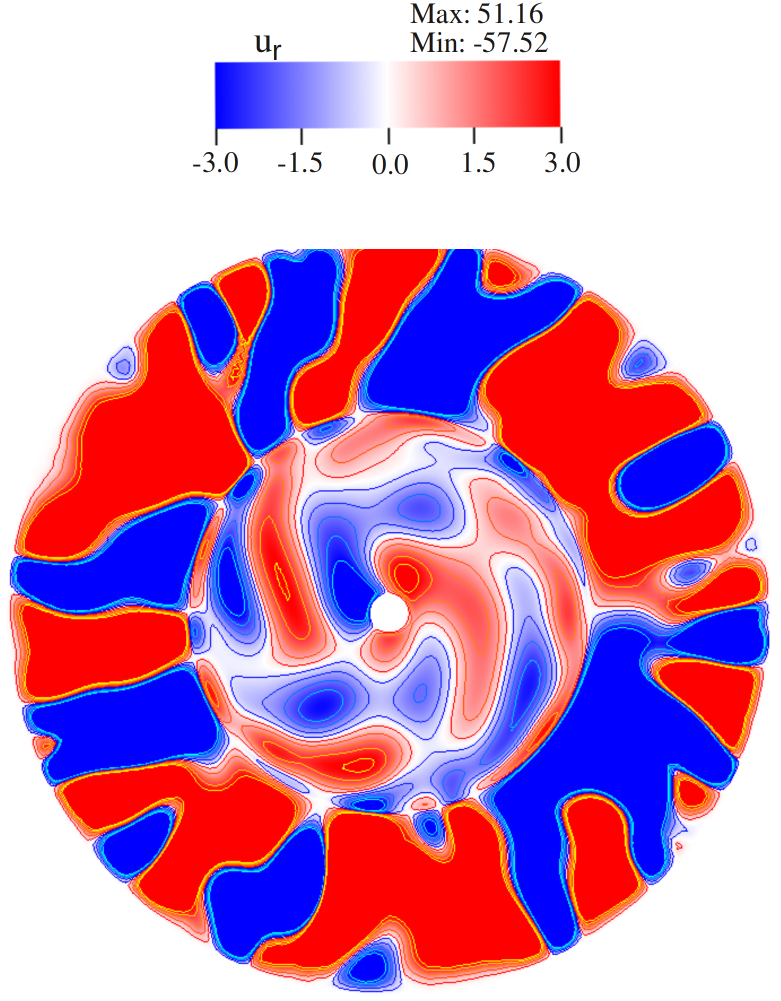} }

\subfloat[]{\includegraphics[angle =0, width=0.3\textwidth]{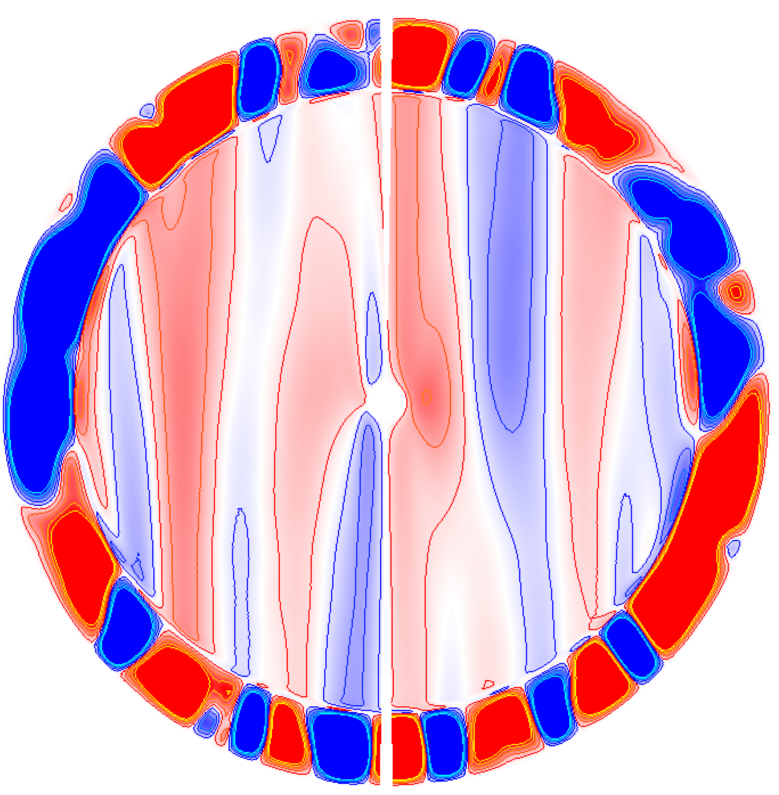} }
\subfloat[]{\includegraphics[angle =0, width=0.3\textwidth]{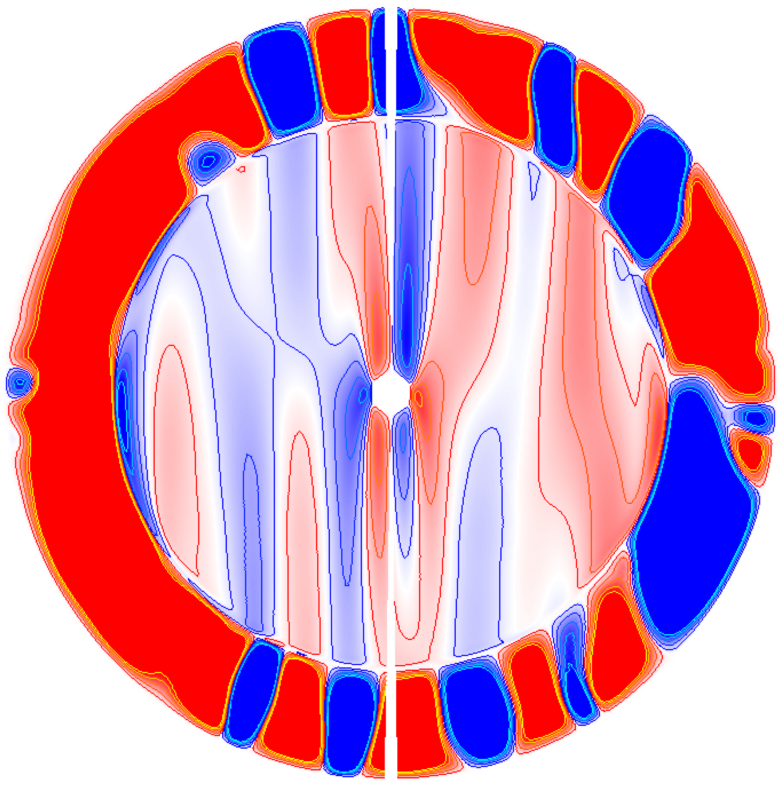} }
\subfloat[]{\includegraphics[angle =0, width=0.3\textwidth]{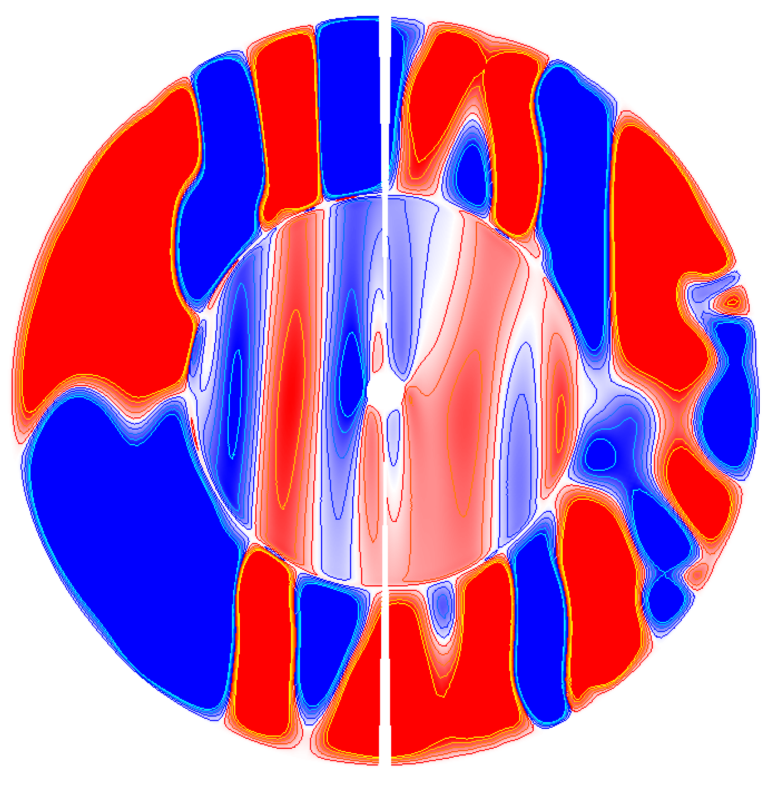} }
\caption{ Equatorial ($u_r$) and azimuthal ($u_\phi$) sections 
for 
$D~(\delta_{o/c+o}=0.2)$,
$\delta\_0.3$, $\delta\_0.5$ in (a),\,(b) and (c)
respectively.
}

\label{fig:fig12}
\end{center}
\end{figure}
It can also be seen that the MO tends to form columnar flows which is a distinctly different feature. Inspite of the fact that core convects strongly, the energy in the MO dominates over that of the core. 
We found that in those cases with small $\nu_{o/c}$, the pattern of flow interaction between the layers is quite different. Both the layers show extremely turbulent flows. The radial velocity range of such system ($\nu_{o/c}=5$) is also the highest among all the cases studied so far. However, as far as the velocity ratio ($U_{o/c}$) is concerned, the change does not follow any pattern with reference to the corresponding change in $\nu_{o/c}$. For example, $U_{o/c}$  drops to 3.14 from 3.2 when $\nu_{o/c}$ drops to 15 from 25. But $U_{o/c}$ again rises to 3.23 with further drop in $\nu_{o/c}$ to 5.
The azimuthal section Fig.\ref{fig:fig10}(d) shows that the effect of rotation in the MO is more prominent, relative to that found in earlier cases (Figs.\ref{fig:fig3}a,b,c, \ref{fig:fig4}a,b,c, \& \ref{fig:fig7}a,b,c). 
Near the equator, presence of shearing flows can be noted.
Radial sections show that the number of columns increases at lower viscosity ratio. These narrower columns do not have sharp boundaries and they remain interconnected (Fig.\ref{fig:fig11}). 
\begin{figure} [!htb]
 \begin{center}
\includegraphics[angle =0, width=0.5\textwidth]{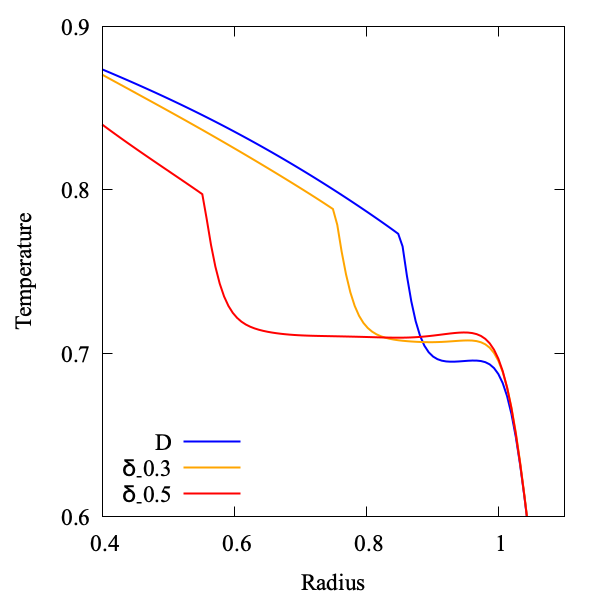} 
\caption{ \small {Snapshot of temperature distribution across model for varying Magma ocean thickness. 
}}
\label{fig:fig13}
 \end{center}
\end{figure}
\subsection{Effect of change in Magma ocean thickness ($\delta_{o/{c+o}}$)}

Depending on the extent of melting due to Earth's collision with other planetary bodies, the thickness of the MO could vary. It also might be possible to melt a large volume of mantle.
Thus, we performed separate simulations for thickness ratios of the two layers ranging from 0.2 to 0.5. 
Increased thickness of the MO does not affect the flow in the core as much as it affects the ocean flow (Fig.\ref{fig:fig12}). Equatorial sections in Fig.\ref{fig:fig12} show that as the thickness of the MO increases, the scale of the flow also increases (Fig.\ref{fig:fig12}c), and the total number of cells drops. The corresponding azimuthal section (Fig.\ref{fig:fig12}f) reveals that flow in a thicker MO, whilst already typically columnar, shows further susceptibility to the rotational constraint within the layer.

If we examine the radial temperature profile (Fig.\ \ref{fig:fig13}), it is seen that the thickness does not have a strong effect on the overall temperature. Being well mixed in the centre of the layer, the changes near the boundaries of the MO are similar for each case.

\section{Conclusions}
\begin{figure}[!h]
\begin{center}
\includegraphics[width=140mm]{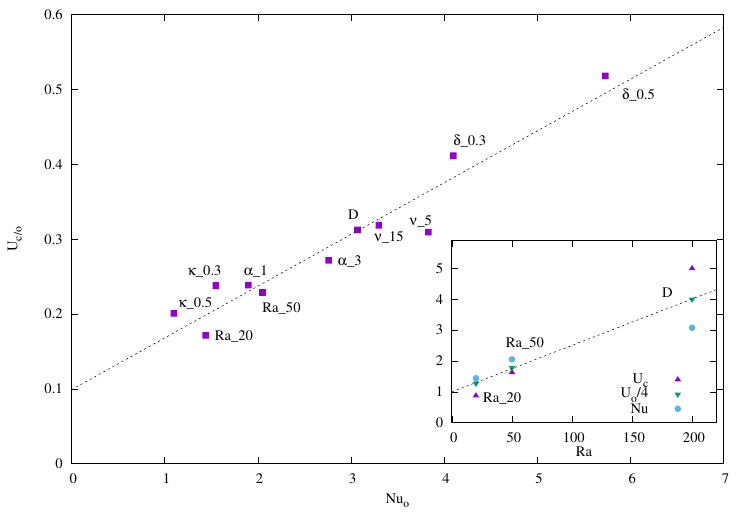}
\caption{
Relationship between $\Nu_o$ and flow in the core driven by coupling with the MO, $U_{c/o}=u_c^{RMS}/u_o^{RMS}=1/U_{o/c}$.
The simulations have been given equal weight for the best-fit 
$U_{c/o}=0.069\,\Nu_o + 0.099$\,.
({\it Inset}) $U_c$, $U_o$ and $\Nu$ as a function of $Ra$.  The fit for $U_o$ is $U_o/4=0.015\,Ra+1.0$\,.
}
\label{fig:fig14}
\end{center}
\end{figure}
The possibility of the presence of a Magma ocean (MO) in planetary bodies has 
attracted much interest over the last few decades.  The formation of the MO is a natural consequence of the differentiation process in such bodies. For Earth, it is widely believed that a basal MO enveloped the core.
The fact that these two fluid layers had comparable physical and chemical properties, introduces the possibility of significant core-MO coupling,
in terms of the flow structures that may be realised.

We have explored the  flow interaction between the two layers
for a range of parameters ratios $\kappa_{o/c}$, $\alpha_{o/c}$, $\nu_{o/c}$, $\delta_o/(\delta_c+\delta_o)$, with 
a default model based on values available data in literature.
 We have started with Ekman numbers sufficiently 
small to see rotational effects, while the supercriticality of the 
Rayleigh number is strongly affected by the parameter ratios.
Nevertheless, the model consistently shows dominance of the MO flows over the core flow.

Varying the ratio parameters separately, 
we are able to see an interesting range of organised and turbulent
flows, either dominated or weakly permeated by columnar rolls, bands,
and sometimes both (cf.\ \cite{garcia2019polar}). Increasing value of $\kappa_{o/c}$ decreases the intensity of convection in MO 
and causes MO to cool more slowly. 
Whereas higher values of $\alpha_{o/c}$ strengthen both MO and core flows. 
Flow turbulence is promoted at lower $\nu_{o/c}$ and greater $\delta_o/(\delta_c+\delta_o)$.  
In all cases, however, when comparing the flow midway through the 
MO and at the same distance below the core-MO interface, it is clear 
that the MO drives a rotationally-constrained flow the core. Table \ref{table2} lists the respective rise and fall in ${U_{o/c}}$. This measure
shows non-trivial dependence on the ratio parameters, but in 
all cases we have simulated it is $O(1$--$10)$. Fig.\ref{fig:fig14} summarises all case studies for current work. $\Nu_o$ is obviously affected by the parameters, but flow in the MO can easily drive a flow of similar magnitude in the core. Values of $U_c$, $U_o$ and $\Nu$ show increasing trend with the increase in $Ra$ (inset). When traced, $\kappa_{o/c}$ cases show moderate, slightly non-linear slope of declination with respect to the super-criticality of the system (Fig.\ref{fig:fig14}). On the other hand, both $\alpha_{o/c}$ and $\delta_o/(\delta_c+\delta_o)$ show a positive correlation with increasing $\Nu_o$. Interestingly, $\nu_{o/c}$ cases seem to maintain $U_{c/o}$ even if $\Nu_o$ changes.

Some studies \citep{ziegler2013implications, stixrude2020silicate} have assumed that the presence of basal MO will not only suppress convection in core, 
but that the core flow is only toroidal. 
We have shown this suppression explicitly, but find that the flow is far from purely toroidal.
\cite{stixrude2020silicate} implemented the mixing length theory which was originally proposed for planetary core dynamos \citep{curtis1986magnetostrophic, christensen2010dynamo} and showed that magma ocean flow is poloidal and it can generate magnetic field.  Our calculations also supports the poloidal flow for the MO,
 and additionally, we found that both the poloidal and toroidal components in core are comparable as  the radial component of the core flow is comparable to its RMS value. This could have significant implications for a geodynamo, opening the 
possibility of a dynamo generated in the core, where the core flow 
itself is driven by a convecting MO. This motivates the investigation of full magnetohydrodynamic studies for this configuration. Adding a magnetic field may change the scale of flow and its vertical structure. We have done simulations at moderate $Ek$ and it cannot be ruled out that the $Nu$-$Ra$ relation changes at lower $Ek$ as it does in rotating Rayleigh- Bernard convection i.e. trending towards $Nu\sim Ra^{3/2}$ or $Nu\sim Ra^3$.


\section{Acknowledgements}

The simulations for this work was done under the Inspire Faculty project awarded to UD by the Department of Science and Technology (DST), Government of India [DST/INSPIRE/04/2016/001582]. UD also acknowledges Council of Scientific and Industrial Research (CSIR), Government of India [25WS(0009)/2023-24/EMR-II/ASPIRE] and Patna University [01/R\&DC/RP/PU/Sanction] for their support at the final stage of this work.
CJD acknowledges support from the Natural Environment Research Council, grant NE/V010867/1.


\bibliographystyle{elsarticle-harv} 

\bibliography{MO}
\end{document}